\documentclass{aa}  
\usepackage{amsmath}
\usepackage{multirow}
\usepackage{graphicx}
\usepackage{txfonts}
\usepackage{hyperref}
\hypersetup{colorlinks,citecolor=blue,linkcolor=blue}
\usepackage{linenoaa}

\begin{document}

\title{On stellar hotspots due to star-planet magnetic interactions}

   \subtitle{How much power can actually be transmitted to the chromosphere?}




\author{Arghyadeep Paul\inst{1}, Antoine Strugarek\inst{1}, Victor Réville\inst{2}
          }

   \institute{Université Paris Cité, Université Paris-Saclay, CEA, CNRS, AIM, F-91191, Gif-sur-Yvette, France\\
              \email{arghyadeepp@gmail.com / arghyadeep.paul@cea.fr}
         \and
             IRAP, Université Toulouse III - Paul Sabatier, CNRS, CNES, Toulouse, France
             }


 
  \abstract
   {Star-Planet Magnetic Interactions (SPMI) have been proposed as a mechanism for generating stellar hot-spots with energy outputs on the order of $10^{19-21}$ watts. This interaction is primarily believed to be mediated by Alfvén waves, which are produced by the planetary obstacle and propagate towards the star. The stellar atmosphere, being a highly structured region, dictate where and how much of this incoming energy can actually be deposited as heat.}
   {The stellar transition region separating the chromosphere from the corona of cool stars gives rise to a significant variation of the Alfvén speed over a short distance, and therefore a reflection of the Alfvén waves at the transition region is naturally expected. We aim to characterize the efficiency of energy transfer due to SPMI by quantifying a frequency dependent reflection of the wave energy at the stellar transition region and its transmission to the stellar chromosphere.}
   {Magnetohydrodynamic simulations are employed to model the frequency-dependent propagation of Alfvén waves through a realistic background stellar wind profile. The transmission efficiency as a function of the wave frequency is quantified, and further analysis is conducted to characterize the overall energy transfer efficiency of SPMI in several candidate systems where chromospheric hotspots have been tentatively detected.}
   {Low-frequency waves experience greater reflection compared to high-frequency waves, resulting in reduced energy transfer efficiency for lower frequencies. Conversely, the parametric decay instability of Alfvén waves substantially diminishes the energy transfer efficiency at higher frequencies. As a result, there exists a frequency range where energy transfer is most efficient. A significant fraction of the Alfvén wave energy is reflected at the stellar transition region, and in most realistic scenarios, the transmission efficiency to the chromosphere is found to be approximately 10\%.}
   {}


   \maketitle
%

\section{Introduction}
Exoplanets in general can undergo a complex array of interactions with their host stars \citep{Vidotto_2019, Strugarek_2024}. Interactions between an exoplanet's outer layers and the stellar radiation can significantly affect its atmosphere, leading to atmospheric heating and evaporation \citep{Antonio_2023}. Additionally, interactions with the ambient stellar wind can reshape the magnetic field configuration in the vicinity of the planet to form either intrinsic magnetospheres (for magnetized planets) or imposed magnetospheres (for non magnetized planets), analogous to the planets within our own solar system\citep{Strugarek_2018}. Tidal interactions between the planet and its host star may drive orbital migration for the planet \citep{Wu_2024} and, in some cases, even result in a noticeable increase in the star's angular momentum \citep{Penev_2016}. Another intriguing class of interactions, particularly prevalent in close-in planets, is known as Star-Planet Magnetic Interactions (SPMI). In these types of interactions, the coupling between the stellar and exoplanetary magnetic fields forms a magnetic tether, which manifests as magnetic flux tubes connecting the planet to its host star \citep{Strugarek_2018, Fischer_2022}. Such interactions are exceptional in close-in exoplanets due to the presence of an Alfvén surface around each star, which is a three-dimensional boundary where the accelerating stellar wind speed matches the local Alfvén speed \citep{Strugarek_2022, Vidotto_2023}. Only planets located within this Alfvén surface can transmit any form of influence back towards the host star. Although all planets within the solar system orbit in a super-Alfvénic wind, analogous sub-Alfvénic interactions are observed within a planet's magnetosphere between natural satellites and their host planets (e.g., the interactions between Io/Europa/Ganymede, and Jupiter)\citep{Saur_2013}.

\begin{figure*}
    \centering
    \includegraphics[width=1\linewidth]{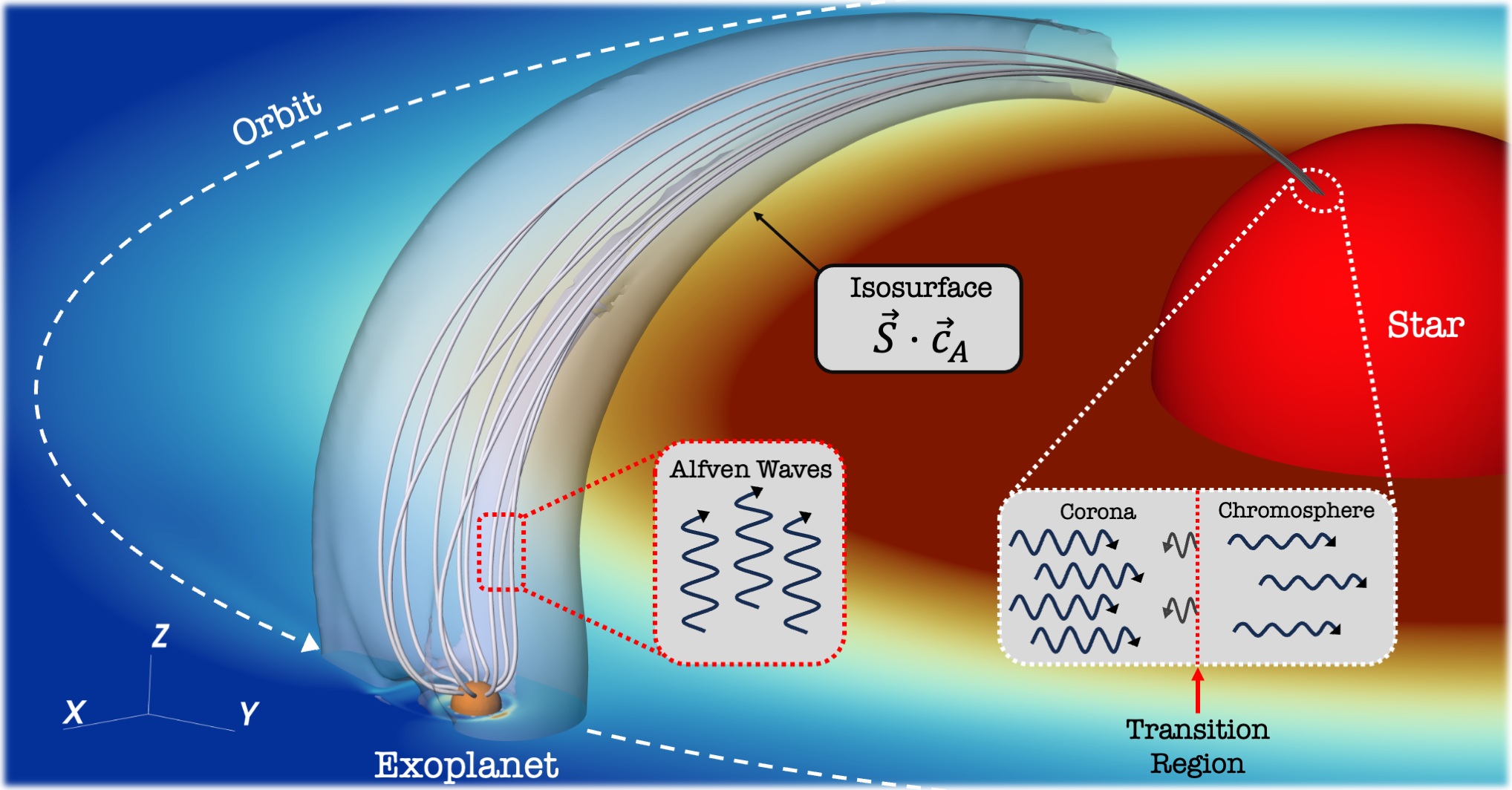}
    \caption{A schematic illustrating key elements of Alfvén wave-mediated star-planet magnetic interactions. The funnel-shaped isosurface represents an isocontour of the Poynting vector projected along the Alfvén characteristics. Magnetic field lines connect the planet to the star, enabling energy transfer from the planet to the star, with a portion reflected at the stellar transition region and the rest transmitted.}
    \label{fig:schematic}
\end{figure*}

\citet{Shkolnik_2005} and \citet{Shkolnik_2008} reported the first tentative detections of SPMI, observing that the stars HD 179949 and $\upsilon$ Andromedae, both hosting hot-Jupiter companions, exhibited chromospheric activity that was synchronized with the orbital period of the exoplanet. Follow-up studies by \citet{Cauley_2019} sought to further support the concept of SPMI, confirming synchronized activity in certain systems and providing estimates of the power emitted by these chromospheric regions. It is important to note, however, that tracers of star-planet interactions observed in stellar activity indicators have exhibited significant variability \citep{Shkolnik_2008} with clear signals detected at certain epochs and no discernible activity at others (e.g. \citealt{Cauley_2018} for the particular case of HD 189733). According to current scientific understanding, close-in exoplanets located within the Alfvén surface of their host stars can generate substantial energy flux through various mechanisms \citep{Strugarek_2018, Saur_2018}. These include magnetic reconnection between the stellar and planetary magnetic fields \citep{Cuntz_2000}, perturbations caused by the planetary obstacle that can produce Alfvén waves \citep{Saur_2013} and dissipation of mutual magnetic stresses on the stellar chromosphere \citep{Lanza_2013}. As these planetary obstacles move through a sub-Alfvénic plasma, they can generate flow cavities in the dominant stellar wind flow, known as Alfvén wings \citep{Fischer_2022, Strugarek_2015, Strugarek_2019}, which are essentially three dimensional structures that harbor magnetic field lines connecting the planetary obstacle to the star. The white translucent isosurface of ($\textbf{\textit{S}}\cdot \textbf{\textit{c}}_A$) in figure \ref{fig:schematic} shows a pictorial representation of the Alfvén wings generated by an exoplanet within a 3 dimensional simulation (\citealt{Strugarek_2016}, Paul et. al. in prep). The energy flux generated near the planet can travel toward the star along the magnetic field lines within the Alfvén wings in the form of Alfvén waves, thereby establishing a two-way magnetic connection between the planet and the star\citep{Lanza_2012, Lanza_2013, Saur_2013, Cauley_2018, Cauley_2019,Strugarek_2015,Strugarek_2016,Strugarek_2018,Strugarek_2019}. Upon reaching the stellar corona, these Alfvén waves traverse the stellar transition region and eventually reach the chromosphere, where they can dissipate and give rise to chromospheric hotspots. In principle, these hotspots would migrate longitudinally across the surface of the star, synchronized with the orbital motion of the planet \citep{Shkolnik_2003, Shkolnik_2008,Cauley_2018, Cauley_2019, CastroGonzalez_2024}. \citet{Fischer_2019} explored the potential role of SPMI in triggering stellar flares within the system but found no conclusive evidence supporting such a correlation. Similarly, \citet{Ilin_2022} concluded that significantly longer observation times are needed to determine whether the flaring activity in the AU Mic system aligns with the expected signatures of SPMI-induced flares. Building on this, \citet{Ilin_2024} analyzed observations from a sample of over 1,800 exoplanet candidates and identified flares from HIP 67522 as the most promising case for SPMI-triggered stellar flares. Despite numerous studies investigating the link between star-planet magnetic interactions and stellar flares, the results to date remain far from definitive \citep{Klein_2022, Loyd_2023}.

Much of the observational evidence supporting theories of sub-Alfvénic Star-Planet Magnetic Interactions (SPMI) is derived as physics driven analogies from solar system planets with sub-Alfvénic conditions within their magnetospheres. This includes Earth's artificial satellites \citep{Drell_1965} and Jupiter with its natural satellites. Energetic particle populations have indeed been observed within the footprint of the Alfvén wing associated with Io on Jupiter's surface \citep{Clark_2020}. Hotspots in ultraviolet wavelengths have also been distinctly observed in Jupiter's polar regions, corresponding to the magnetic footprints of Io, Ganymede, and Europa \citep{Clarke_2002}. Auroral radio emissions from Jupiter and other outer planets also act as indicators of the magnetic field strength around the planet \citep{Nichols_2022}. Detection of such emissions in exoplanet populations could provide valuable insights into the extent of SPMI within these systems and help constrain the magnetic field strengths of the exoplanets. However, no definitive detections of auroral radio emissions from exoplanets have been reported to date \citep{Zarka_2018,Vedantham_2020,Pineda_2023,Shiohira_2023}. In the context of detecting chromospheric hotspots related to SPMI, numerous attempts have been documented in the literature, with varying degrees of robustness. Notable studies include those by \citet{Shkolnik_2003}, \citet{Shkolnik_2005}, \citet{Gurdermir_2012}, \citet{Shkolnik_2008}, \citet{Cauley_2018}, and \citet{Cauley_2019}. In particular, \citet{Cauley_2019} was the first to present flux-calibrated absolute values of the power associated with SPMI observations across four different exoplanetary targets, estimating a flux in the range of $10^{20}$ to $10^{21}$ W. Among the various scaling laws commonly used to estimate the power generated by an obstacle in a magnetized plasma, it has been found that, under realistic stellar and planetary conditions, the energy released purely by the process of magnetic reconnection between the planetary and stellar magnetic fields can account for power on the order of $10^{18}$ W \citep{Lanza_2009}. \citet{Saur_2013} provided scaling laws for a scenario where SPMI power is generated by Alfvén wings associated with the planetary body, with the energy budget estimated to reach approximately $10^{19}$ W for realistic systems considered in their study. These scaling laws were validated against sub-Alfvénic satellite-planet interactions observed in the solar system. Although the magnetic reconnection scenario considerably underestimates the observed power in SPMI, the second scenario, which involves Poynting flux generation by Alfvén wings, only barely approaches the observed SPMI power levels, leaving little margin to account for powers exceeding $10^{20}$ W. Consequently, \citet{Cauley_2019} based their power estimations on the model proposed by \citet{Lanza_2013}, which attributes the power budget to magnetic stresses induced by the planet's orbital motion through the stellar magnetic field. Though yet to be fully validated by numerical simulations, the scaling law proposed by \citet{Lanza_2013} is currently one of the few models that can reasonably account for SPMI power outputs in the range of $10^{20}$ to $10^{21}$ W for hot-Jupiter systems with realistic parameters. 

Numerical simulations have proven to be an essential tool for constraining various parameters associated with SPMI in close-in exoplanetary systems. \citet{Matsakos_2015} classified the morphology of SPMI in close-in hot Jupiter systems, identifying four general types of interactions that depend on a combination of fundamental stellar, planetary, and orbital parameters. \citet{Strugarek_2016} developed numerically motivated scaling laws to estimate the magnetic torque acting on exoplanets and the energy flux produced by SPMI for different planetary magnetic field orientations. They also found reasonable agreement between the Poynting flux generated in self-consistent 3D MHD simulations and the analytical scaling laws proposed by \citet{Saur_2013}.  \citet{Strugarek_2022} further evaluated the power budget of SPMI for a particular exoplanetary system and predicted its temporal modulation to facilitate comparison with observational data. Their findings indicated that, while the Alfvén wing model estimates the generated power with reasonable accuracy, the power output remains insufficient to account for the full energy budget of the observed signal. \citet{Fischer_2022} analyzed the interactions between Alfvén wing structures in scenarios where multiple planets generate their respective Alfvén wings, which subsequently interact with one another. They observed a notable intensification of the resulting Poynting flux, particularly during the initial phases of the wing-wing interaction. As the scientific community continues to refine these models and extend them to interpret potential observations of SPMI in known exoplanetary systems, several uncertainties and unknowns remain. These challenges primarily revolve around the exact mechanisms responsible for driving the observed energy outputs, as well as efficiencies in the processes of energy generation, transfer, and emission. This paper specifically focuses on addressing the latter category of unknowns, particularly the efficiency factors associated with SPMI. Briefly stated, there are several efficiency factors to be considered during the stages of energy transfer in systems exhibiting SPMI. First, near the location of the exoplanet, there exists an efficiency factor related to the conversion of the stellar Poynting flux at the planet's orbital position into the Poynting flux directed back toward the star along the connecting magnetic field lines shown in figure \ref{fig:schematic}. This efficiency has been the focus of many SPMI models so far \citep{Saur_2013, Lanza_2013}. Secondly, the wave structures carrying energy along these magnetic field lines may dissipate during their propagation due to interactions with the ambient medium and its inhomogeneities giving rise to another efficiency factor. Thirdly, as these waves approach the star, they encounter the stellar transition region, a sharp discontinuity in plasma properties. As highlighted in figure \ref{fig:schematic}, this discontinuity introduces another efficiency factor, as a portion of the wave energy is likely reflected back by the transition region, much like the reflection process envisioned by \citet{Leroy_1979}, \citet{Leroy_1980} and \citet{Leroy_1981} in the context of solar-driven Alfvén waves. Finally, of the energy that successfully reaches the stellar chromosphere, an additional efficiency factor governs how effectively this energy is dissipated into detectable emissions via various emission mechanisms. Efficiency factors like these are pivotal for estimating the energy budget in SPMI systems based on observable emission signatures, making their accurate determination crucial. As an initial step, this paper focuses on estimating the efficiency of energy transfer via Alfvén waves as they traverse the stellar transition region - a critical factor in the chain of energy generation and its eventual dissipation and emission within SPMI exhibiting systems. 

The paper is organized as follows. Section \ref{sec:numer_set} describes the details of the numerical setup used to perform the study. Section \ref{sec:results} highlight the main results obtained in this study with a quantification of the efficiencies involved in transmission of SPMI powers through the stellar transition region. Section \ref{sec:observations_SPMI detect} highlights the implications of the obtained results for the observational detections of SPMI and finally section \ref{sec:conclusions} summarizes the paper and presents some concluding remarks.  

\section{Numerical setup \label{sec:numer_set}}
As illustrated in the schematic depicted in \ref{fig:schematic}, a close-in exoplanetary system typically consists of a host star and its orbiting exoplanet, linked by magnetic field lines. Alfvén wings, depicted as the translucent isosurface in the schematic, form due to the interaction between the star and the planet. These Alfvén wings channel a significant amount of Poynting flux back toward the star in the form of Alfvén waves. The Alfvén waves travel along the magnetic field lines within these wings, connecting the exoplanet to the star, as shown in the schematic. In simple terms, the one-dimensional numerical domain used in this study, as explained in the subsequent sections, corresponds to a computational grid along a single magnetic field line connecting the exoplanet to the star. Within this numerical domain, we first describe the stellar wind properties, as detailed in the following section.

\subsection{Background solar wind}
 We adapt the numerical solar wind setup developed by 
\cite{Reville_2018} that involves solving the following set of ideal MHD equations in one dimension using the PLUTO code:
\begin{equation}\label{eq:MHD_eqs}
\begin{aligned}
    \frac{\partial \rho}{\partial t} + \nabla \cdot(\rho \mathbf{v})  &=  0 \\
    \frac{\partial (\rho  \mathbf{v})}{\partial t} + \nabla \cdot \left[\rho \mathbf{v}\mathbf{v} - \mathbf{B}\mathbf{B}\right] + \nabla \left( p + \frac{\mathbf{B}^2}{2} \right)  &= -\rho \nabla \Phi \\
    \frac{\partial \mathbf{B}}{\partial t} + \nabla \times (c\mathbf{E}) &= 0 \\
    \frac{\partial E_{t}}{\partial t} + \nabla \cdot \left[ \left( \frac{\rho\mathbf{v}^2}{2} +\rho e + p\right)\mathbf{v}  + c\mathbf{E}\times \mathbf{B} \right] &= Q
\end{aligned}
\end{equation}
where $\rho$ is the mass density, $\mathbf{v}$ is the gas velocity, $p$ is the thermal pressure and $\mathbf{B}$ is the magnetic field. The vector fields are defined in spherical coordinates ($r, \theta, \phi$) with variations only in the radial direction. Such a prescription using the above equation describes the evolution of a single radial flux tube under the influence of a gravitational potential given as:
\begin{equation}
    \Phi = -\frac{GM_{\star}}{r}
\end{equation}

A factor of 1/$\sqrt{4\pi}$ has been absorbed in the definition of $\mathbf{B}$. $E_{t}$ is the total energy density which can be described as:
\begin{equation}\label{eq:energy}
    E_{t} = \rho e + \frac{\rho \mathbf{v}^{2}}{2} + \frac{\mathbf{B}^2}{2}
\end{equation}

The quantity $Q$ in the RHS of the energy equation is a source term comprising of three components
\begin{equation}\label{eq:RHS_energy}
    Q = Q_h -Q_r -Q_c
\end{equation} 
 that describe the usual heating, cooling and thermal conduction terms for a typical wind of a Sun-like star. The individual terms are implemented as follows:
 \begin{equation}
     Q_h = \frac{F_h}{h} \left(\frac{R_{\star}}{r}\right)^2 exp \left[- \left (\frac{r- R_\star}{H}\right)\right]
 \end{equation}
where $H = 1 R_\star$ is the heating scale height, and $F_h = 1.5 \times 10^5 $erg cm$^{-2}$s$^{-1}$ is the stellar photospheric energy flux. This value of the heating rate has been chosen in order to obtain a mass loss rate of $\dot{M} = 3 \times 10^{-14} M_\odot yr^{-1}$, that is consistent with observations for a Sun-like star \citep{Reville_2018}.  The radiation term describes an optically thin radiative cooling prescribed as :
\begin{equation}
    Q_r = n^2 \Lambda(T)
\end{equation}
with n and T being the electron density and temperature respectively. The function $\Lambda(T)$ is described in \cite{Athay_1986}. The thermal conduction flux comprises of a combination of a collisional and collisionless prescription:
\begin{equation}
    Q_c = \nabla \cdot (\alpha \mathbf{q_s} + (1-\alpha)\mathbf{q_p})
\end{equation}
 where \textbf{$q_s$} is the usual Spitzer-Harm conduction with the value of $\kappa_0 = 9 \times 10^{-7}$ in cgs units. $\mathbf{q_p}$ is the free-stream heat flux as defined in \cite{Hollweg_1986} and is given by $\mathbf{q_p} = 3/2 p\textbf{v}$. The coefficient $\alpha$ is defined as:
 \begin{equation}
     \alpha = \frac{1}{\left[ 1 + \left(\frac{r- R_{\star}}{r_{coll}- R_{\star}}\right)^4\right]}
 \end{equation}
 This prescription creates a smooth transition between the collisional and the collisionless regimes at a characteristic height of $r_{coll} = 5 R_\star$.
An ideal equation of state provides the closure as $\rho e = p/ (\gamma -1)$ wherein, $\gamma$ is the ratio of specific heats having a value of 5/3. The flux computations have been performed with the second order accurate variation of the Harten-Lax-vanLeer (HLLD) solver and the solenoidal constraint ($\nabla\cdot\textbf{B}$ = 0) is imposed by coupling the induction equation to a  Generalized Lagrange Multiplier (GLM) and solving a modified set of conservation laws in a cell-centered approach \citep{Dedner_2002}.

The 1D computational grid extends from the stellar photosphere assumed to be located at 1$R_\star$ up to a radius of 20$R_\star$. To save on computational time, we first initiate a steady state background stellar wind solution on a computational grid that is divided as follows. The region from 1$R_\star$ to 1.001$R_\star$ is discretized into 256 uniformly spaced grid cells. Thereafter, the region from 1.001$R_\star$ to 1.5$R_\star$ is divided into 4096 grid cells having a `stretched' configuration. This imposes that the grid cells become recursively larger by a constant geometrical factor with increasing radial distance. Finally the region from 1.5$R_\star$ to 20$R_\star$ is discretized into 28416 uniform grid cells. Once the background stellar wind reaches a stationary solution in the aforementioned grid, we change the grid layout into a higher resolution one. The new grid has a similar prescription till 1.001$R_\star$, but thereafter, the region between 1.001 and 1.049 is divided into 2112 stretched grid cells. Further, from 1.049$R_\star$ to 4.1$R_\star$, we employ 48064 uniform grid cells. Finally, the region from 4.1$R_\star$ to 20.0 $R_\star$ is then covered by 64 stretched grid cells. This layout was motivated by the aim of the study which was to investigate Alfv\'en waves traveling from an exoplanet situated at 4$R_\star$ towards its host star. As such, we resolve with a very high grid cell count, the region between the stellar photosphere and the planet. Anything beyond the planet is less critical for this study and therefore has been simply buffered with a relatively coarse grid. We apply a zero-gradient boundary condition to waves traveling towards the inner boundary, while maintaining all other physical quantities at their equilibrium values. Conversely, at the outer boundary, a zero-gradient boundary condition is imposed universally on all physical quantities. As the wind itself is supersonic and super-Alfv\'enic at the outer boundary, no wave reflection is physically possible.

\begin{figure*}
    \centering
    \includegraphics[width=1\linewidth]{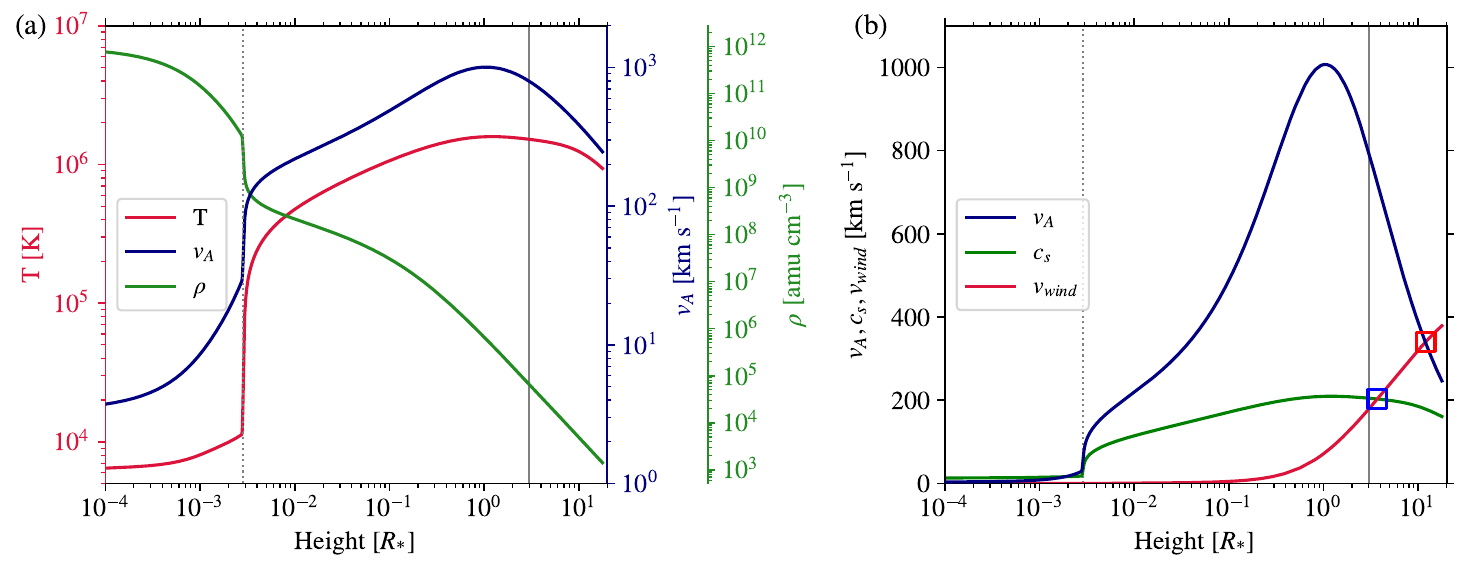}
    \caption{Panel (a) shows the temperature, Alfv\'en speed and density of the background stellar wind model on the highest resolution grid whereas panel (b) highlights a comparison of the Alfv\'en speed, sound speed and bulk radial speed profiles. The grey dotted and solid lines in both panels represent the location of the transition region and the planet respectively.}
    \label{fig:background}
\end{figure*}

Figure \ref{fig:background} shows a few physical parameters of the steady state background field solution on the highest resolution grid. In panel (a), the red, green and blue profiles represent the temperature, density and the Alfv\'en speed profiles of the background stellar wind on their individual ordinates. The lower boundary of the domain behaves as a stratified atmosphere fixed at a specific temperature of T $\sim$ 6000 K. The prescribed phenomenological heating term in equation \ref{eq:RHS_energy}, $Q_h$, heats up the stellar atmosphere up to a maximum temperature of $\sim$ 1.7 $\times$ 10$^{6}$ K as seen by the temperature profile denoted by the red curve in panel (a) figure \ref{fig:background}. Such a prescription also leads to the formation of a sudden temperature jump, also known as the transition region (TR), at a height of $\sim$ 2.8 $\times$ 10$^{-3}$ R$_\star$ highlighted by the grey dotted line in panel (a) of figure \ref{fig:background}. Panel (b) of figure \ref{fig:background} shows a clear comparison of the Alfv\'en speed, the sound speed and the bulk stellar wind speed within the domain on a linear ordinate. Upon comparison, it can be clearly seen that the stellar wind turns super-sonic at a height of $\sim$ 3 $R_{\star}$ (orbital radius of $\sim$ 4 $R_{\star}$ from the center of the Sun) and super-Alfv\'enic at a height of $\sim$ 13 $R_{\star}$. The blue and red squares represent the locations where the wind speed equals the local sound speed and the Alfv\'en speed respectively.

\subsection{Exoplanet characteristics}
We flag two grid cells located at 4$R_\star$ (height of 3$R_\star$ from the stellar surface) to simulate a close-in hot Jupiter-like planet that injects circularly polarized, inward propagating (towards the star) Alfv\'en waves into the domain. As the domain is primarily one dimensional, the region between 1$R_\star$ to 4$R_\star$ can be considered to be along one of the magnetic field lines that connects the exoplanet to the star in figure \ref{fig:schematic}. The location of the planet in the 1D domain is marked with solid grey vertical line in panels (a) and (b) of figure \ref{fig:background}.  It can be clearly seen from panel (b) that at the location of the planet, the stellar wind is both sub-sonic as well as sub-Alfv\'enic. At the injecting grid, the pressure, density, $v_r$ and $B_r$ are untouched whereas the quantities $v_\theta$, $v_\phi$, $B_\theta$, $b_\phi$ (together denoted as $v_\perp$ and $b_\perp$) are varied with time in order to inject inward propagating Alfv\'en waves into the domain. To that end, we define the El$\rm\ddot{s}$asser variables as:
\begin{equation}
    \mathbf{z^\pm} = \mathbf{v_\perp} \pm \frac{\mathbf{b_\perp}}{\sqrt{\mu_0 \rho}}
\end{equation}
which denote propagating Alfv\'enic fluctuations. For inward propagating Alfv\'en waves, we use the lower `+' (plus) sign in the above equation and define the perturbations as:
\begin{align}
    z^+ = 2|\delta v|\left[ \cos(\omega_0 t) \hat{\theta} + \sin(\omega_0 t) \hat{\phi}\right]
\end{align}
where $\hat{\theta}$ and $\hat{\phi}$ are the unit vectors along their usual directions, $\omega_0 = 2\pi f_0$ is the angular frequency of the wave and the quantity $|\delta v|$ sets the magnitude of the perturbations. For the purposes of this study, we set the value of $\delta v$= 50 km $\rm{s^{-1}}$ with the motivation that the magnetic field perturbations produced by such an amplitude is approximately $\sim 5\times 10^{-2}$ times the local background magnetic field. This puts the simulations within the limit of small-amplitude perturbations. It is imperative to achieve an appropriate balance between the perturbation amplitudes considered and the spatial resolution used in this study. It is well understood that higher amplitude perturbations can result in the formation of additional shocks and discontinuities in the medium, necessitating significantly higher spatial resolution to capture and evolve these effects accurately \citep{Alielden_2022}. Consequently, another indirect motivation behind the choice of the perturbation amplitude was to remain safely within a range that avoids such potential complications, thereby ensuring overall numerical stability within the model with the resolution considered. A brief outlook on the variation of the perturbation amplitude on the principal results of this study is presented in \ref{sec:tc_sim}. In the analysis that follows, the time is normalized to the Alfv\'en crossing time ($t_A$) for inward propagating waves from the planet to the star and is calculated as:
\begin{equation}
t_A = \int_{4R_\odot}^{R_\odot} \frac{dr}{v_A - v_{wind}} = 2994 s \approx 50 \; \rm min
\end{equation}
 
In the following paragraphs, we elaborate the motivation behind the range of the Alfv\'en wave frequencies we choose to probe in this study. It is reasonable to infer that Alfv\'en waves are triggered by the interaction between stellar magnetic field lines and obstacles within the solar wind plasma. These obstacles could be the planetary body itself for an unmagnetized planet, or the magnetosphere for a planet with an intrinsic magnetic field \citep{Zarka_2007,Saur_2013}. 

Within a simplistic model of encounter, let us assume that a magnetic field line sweeps across the diameter of the obstacle. The time of such an encounter would depend on (a) the size of the obstacle and (b) the orbital speed of the obstacle relative to the magnetic field lines. For a planetary obstacle at an orbital radius of 4 $R_\star$ around a Sun-like star, the orbital velocity can be calculated to be $\sim$ 218 km s$^{-1}$. Considering an obstacle comparable to the size of Jupiter, a field line sweeping through its diameter would interact with the obstacle for $\sim$ 320 seconds. Therefore, the minimum frequency of the Alfv\'en waves that could be generated by such an interaction is of the order of 0.001 Hz. We therefore consider this to be the lower limit of our frequencies. Conversely, the upper limit of the frequency has been set with a similar calculation for an obstacle that is approximately half the size of the smallest planet in the solar system, Mercury ($R_{mercury} \sim 0.35 R_E$), leading to a frequency of $\sim$ 0.1 Hz. Within this range of 0.001 Hz to 0.1 Hz, we probe 10 different frequencies that are logarithmically spaced to span these bounds forming a frequency set given in  table \ref{tab:freqs}. For easier reference throughout the paper, these frequencies are also designated as $f_0$ to $f_9$. The third column of table \ref{tab:freqs} also represents the obstacle (either the planet itself for an unmagnetized case or the magnetosphere for a magnetized planet) radius from which, waves of such frequencies can be expected. Armed with the background solar wind to propagate Alfv\'enic fluctuations in and a logically motivated set of frequencies, we proceed further into the study.
\begin{table}
\centering
\begin{tiny}
\caption{\begin{tiny}The set of frequencies probed in this study.\end{tiny}}
    \label{tab:freqs}
\begin{tabular}{ ||c|c|c|| }
\hline
Frequency [Hz]  & Name & Obstacle Size \\
\hline
$1.0 \times 10^{-3}$ & $f_0$ & 1.52 $R_J$ \\
$1.6 \times 10^{-3}$ & $f_1$ & 0.91 $R_J$\\
$2.7 \times 10^{-3}$ & $f_2$ & 0.54 $R_J$ \\
$4.6 \times 10^{-3}$ & $f_3$ & 0.32 $R_J$ \\
$7.7 \times 10^{-3}$ & $f_4$ & 2.21 $R_E$ \\
$1.29 \times 10^{-2}$ & $f_5$ & 1.32 $R_E$ \\
$2.15 \times 10^{-2}$ & $f_6$ & 0.79 $R_E$\\
$3.59 \times 10^{-2}$ & $f_7$ & 0.47 $R_E$ \\
$5.99 \times 10^{-2}$ & $f_8$ & 0.28 $R_E$ \\
$1.0 \times 10^{-1}$ & $f_9$ & 0.17 $R_E$ \\
\hline
\end{tabular}
\tablefoot{\begin{tiny}The second column highlights the nomenclature of frequencies in this study and the third column corresponds to the obstacle sizes they represent. $R_J$ and $R_E$ correspond to the radius of Jupiter and Earth respectively.\end{tiny}}
\end{tiny}
\end{table}

\section{Transmission of Alfv\'en waves into stellar chromospheres\label{sec:results}}
Upon establishment of a steady state background stellar wind with desirable characteristics, Alfv\'en waves are injected into the domain that propagate from the location of the exoplanet's orbit towards the host star. As illustrated by the profile of the Alfv\'en speed in panel (a) of figure \ref{fig:background}, the variation in $v_A$ from the planet's location towards the star is predominantly smooth and gradual, with the exception of the TR where the relative change is abrupt. As expected, Alfv\'en waves are found to propagate with minimal reflection up to the TR, where this sudden change in Alfv\'en speed causes a significant portion of the wave energy to be reflected back, while the remaining energy is transmitted through the TR towards the stellar surface. We show this with the help of a quantity defined as the wave action flux (WAF). In a stationary uniform medium, the energy of a wave-train is generally expressed in terms of the wave energy density. However, for waves travelling in a streaming medium with a variation of the wave velocity along the direction of propagation, the wave energy density along a wavetrain may not be constant even when the total wave energy is conserved. Therefore, a more convenient approach to mathematically describe wave evolution is in terms of a `wave action density' which, in its simplest form is defined as E/$\omega$ where `E' is wave energy density and $\omega$ is the intrinsic frequency of the wave. In the absence of non-linear interactions, the total wave action behaves as a conserved quantity. For Alfv\'en waves of the form described above, the associated wave action flux can be defined conveniently as \citep{Reville_2018, Jacques_1977,Huang_2022} :
\begin{equation}
    S^{\pm} = \rho r^2\frac{\left(v \mp v_A\right)^2}{v_A}\frac{\lvert z^{\pm} \rvert^2}{8} 
\end{equation}
Within the context of this study, the quantities $S^{+}$ and $S^{-}$ denote the WAF for waves traveling towards and away from the star respectively.

\begin{figure}
    \centering
    \includegraphics[width=1\columnwidth]{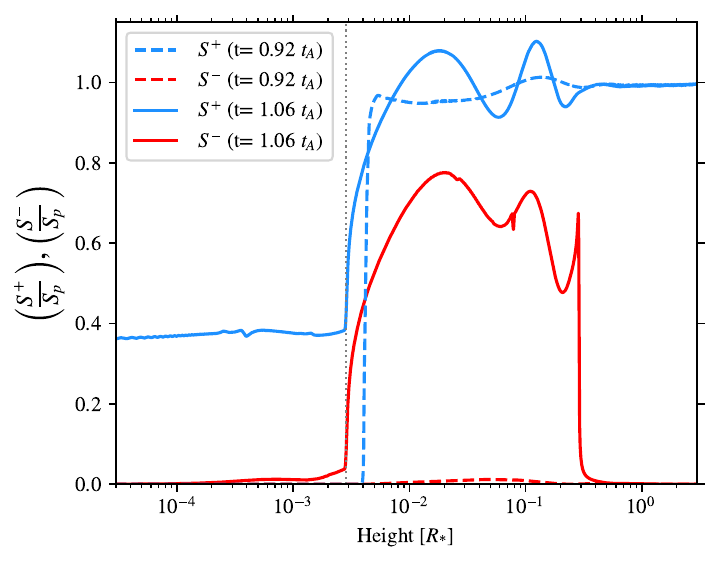}
    \caption{The normalized incoming ($S^+$ in blue) and reflected ($S^-$ in red) WAF components at two distinct times. Dashed lines indicate the values at t= 0.92 $t_A$, before the waves reach the transition region. Solid lines represent the values at t= 1.06 $t_A$, after the waves have reached the stellar boundary. The WAF values are normalized relative to the incoming WAF values at the location of the planet ($S_p$)}
    \label{fig:WAF0.003}
\end{figure}

We monitor the evolution of wave action flux over time as the waves from the planet's location propagate towards the star. For highlighting the general trend in evolution of the WAFs, we explore the profiles at two different times for one of the frequencies considered in the study, specifically, $f_2$. The blue and red dashed lines in figure \ref{fig:WAF0.003} represents the normalized incoming ($S^+$) and reflected ($S^-$) WAF components at t= 0.92 $t_A$, just before the waves reach the transition region (TR). These WAF values are normalized to $S_p$ which is the WAF value at the location of the planet. The figure shows that the incoming component is dominant and the reflected component is negligible before the waves hit the TR. At a later time, t= 1.06 $t_A$, the blue and red solid lines in figure \ref{fig:WAF0.003} represent the normalized WAFs when the waves have reached the stellar boundary. It is evident that a portion of the incoming WAF is transmitted through the TR, while a significant portion is reflected back, as indicated by the much higher $S^-$ value to the right of the TR. This reflected component gradually travels back toward the planet while interacting with the incoming component giving rise to a complex overall evolution pattern. We now explore how the transmission of wave energy varies with frequency.

\begin{figure}
    \centering
    \includegraphics[width=1\columnwidth]{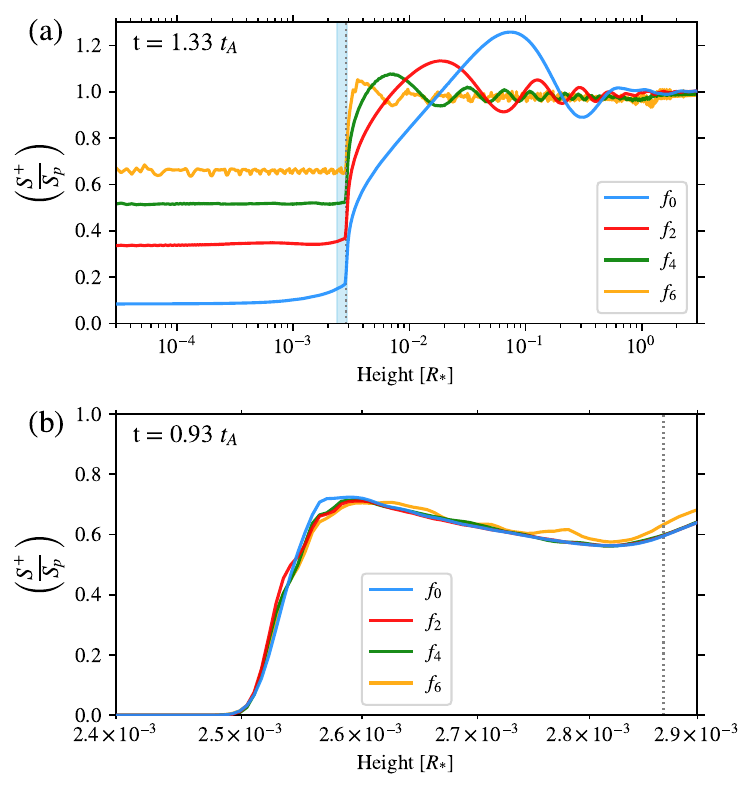}
    \caption{ Panel (a) shows the normalized incoming WAF ($S^+$) profiles for four different frequencies, namely $f_0$, $f_2$, $f_4$ and $f_6$ at t = 1.33 $t_A$ highlighting a frequency dependent transmission of the WAF through the TR. Panel (b) shows a zoomed in view of the WAF for the same frequencies just after the waves cross the TR (t = 0.93 $t_A$). The entire extent of the abscissa in panel (b) is highlighted by the light blue strip in panel (a).}
    \label{fig:diff_freqs_WAF}
\end{figure}

We can see from figure \ref{fig:WAF0.003} that the profile of the normalized incoming WAF ($S^+$) can be conveniently used to derive a transmittance of the WAF, named as $\mathcal{T}_{WAF}$. The transmittance can be defined as the ratio of the WAF values at the surface of the star to that at the location of the planet. Upon normalizing the WAF profiles to $S_p$, $\mathcal{T}_{WAF}$ can simply also be defined as the value of the normalized WAF at the stellar boundary (i.e at r = 1 $R_\star$). As per the foundations laid in the above sections in terms of the Alfv\'en wave properties, we inject wave trains from the planet towards the star with the frequencies given in table \ref{tab:freqs} . The general evolution pattern of the waves of all frequencies show a similar trend as that in figure \ref{fig:WAF0.003}, i.e, the waves travel up to the TR with a minimal amount of reflection, and upon encountering the TR, a significant portion of the WAF is reflected. The reflected waves interact with the incoming waves leading to stationary wave-like patterns to the right of the TR in the WAF profiles. More interestingly, it is seen that there exists a variation in the magnitude of the normalized WAF at the stellar boundary, i.e the transmittance $\mathcal{T}_{WAF}$, for different frequencies. We highlight this in panel (a) of figure \ref{fig:diff_freqs_WAF} with the help of the WAF profiles for four distinct frequencies from the set given in table \ref{tab:freqs}. It is also evident from panel (a) that $\mathcal{T}_{WAF}$ increases with increasing frequency of the waves. 

Panel (b) of figure \ref{fig:diff_freqs_WAF} provides a close-up view of the WAFs near the transition region for the frequencies highlighted in panel (a). While panel (a) displays the WAFs at a stationary state (t = 1.33 $t_A$), panel (b) focuses on an epoch (t = 0.93 $t_A$) when the waves have just crossed the TR. The entire span of the x-axis in panel (b) corresponds to the light blue vertical strip in panel (a), illustrating the extent of the zoom. Panel (b) highlights that a maximum transmission of the WAF, $\sim$ 0.72, occurs at the very leading edge of each wave train and that maximum value is practically independent of the wave frequency. This indicates that the change in the steady state $\mathcal{T}_{WAF}$ as seen in panel (a) of figure \ref{fig:diff_freqs_WAF} arises purely from the interaction of the incident waves with the reflected wave components at the TR. This establishes an upper limit on the $\mathcal{T}_{WAF}$ for the wind profile considered in this study where, $\mathcal{T}_{WAF} \sim$ 0.7 in an ideal scenario.

\subsection{Quantifying the transmittance\label{sec:tc_sim}}
\begin{figure}
    \centering
    \includegraphics[width=1\columnwidth]{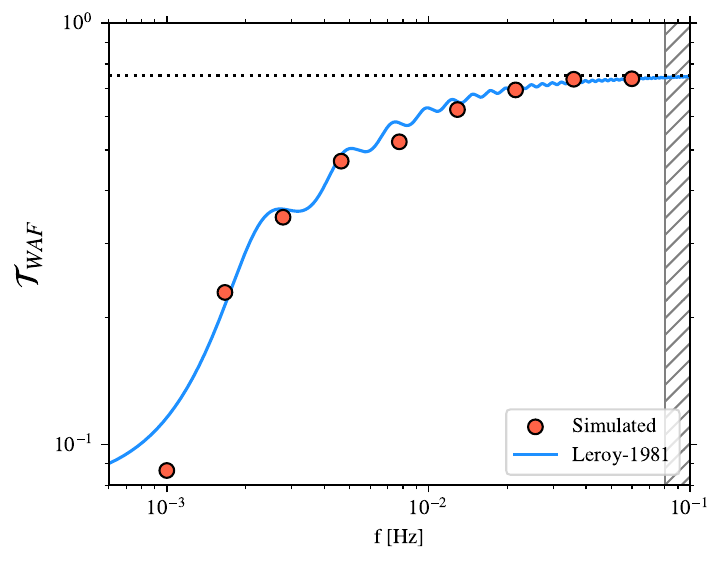}
    \caption{The red scatter points are the transmittance ($\mathcal{T}_{WAF}$) obtained from the simulations of frequencies $f_0$ to $f_8$. The blue solid line is the analytical profile of transmittance obtained using the Leroy-1981 model. The horizontal dotted line represents the asymptotic value of the $\mathcal{T}_{WAF}$ profile having a value of 0.75 obtained using the Leroy-1981 model. The grey hatched region is where PDI was found to significantly influence the transmittance.}
    \label{fig:TC_leroy}
\end{figure}

As elaborated in section \ref{sec:results}, the transmittance of Alfv\'en waves is frequency dependent. Naturally, the next step in the process is to characterize this dependence for the frequency range considered in this study. An approximate stationary state is achieved at the stellar boundary for the WAF profiles of all frequencies after t $\sim$ 1.03 $t_A$. The term "approximate" is used because, as seen from the yellow curve ($f_6$) in panel (a) of Figure \ref{fig:diff_freqs_WAF}, the WAF reaching the stellar (left) boundary exhibits small-scale fluctuations around a certain value in some cases. To tackle this, we determine the transmittance ($\mathcal{T}_{WAF}$) by taking the median of the values obtained at the stellar boundary between t = 1.06 $t_A$ and t = 1.33 $t_A$. We have not calculated $\mathcal{T}_{WAF}$ for $f_9$ at this stage because the Alfvén waves associated with $f_9$ undergo parametric decay, leading to a significant reduction in transmittance. Consequently, this region is shaded with a grey hatched line in Figure \ref{fig:TC_leroy} to indicate that it does not conform to the expected trends from the other simulations and the Leroy-1981 model (see below). The onset of parametric decay instability (PDI) for certain frequencies requires special consideration and is discussed in detail in  \ref{sec:PDI}.

The red scatter points in Figure \ref{fig:TC_leroy} represent the transmittance derived from simulations conducted across the frequency range $f_0$ to $f_8$. Within this frequency spectrum, it is observed that low-frequency waves exhibit significant reflection, with a transmittance of only $\mathcal{T}_{WAF} \sim 0.08$ at a frequency of $10^{-3}$ Hz ($f_0$). As the frequency increases, $\mathcal{T}_{WAF}$ progressively rises and eventually saturates around $\mathcal{T}_{WAF} \sim 0.73$ for higher frequencies, such as $f_7$ and $f_8$. It is important to note that the frequencies in this study are logarithmically spaced, reinforcing the robustness of the saturation trend. As discussed in section \ref{results}, on the basis of panel (b) of Figure \ref{fig:diff_freqs_WAF}, a maximum value of $\mathcal{T}_{WAF} \sim 0.7$ is achievable in the absence of reflected waves at the TR. For the higher frequencies, the reflected WAF trends to be minimal and therefore the observed saturation of the $\mathcal{T}_{WAF} \sim 0.75$ is consistent with the previous assertion.

We now turn our attention briefly to a simplified yet significant variation of the previous scenario, focusing on the emission of an Alfvén wave pulse from the planet instead of a continuous wave train. Limiting the wave injection to a single wavelength reveals that the evolution pattern closely resembles the WAF profiles observed with continuous injection up to the first leading wavelength. However, a steady state is obviously not attained in this case. Given that a maximum value of  of $\mathcal{T}_{WAF} \sim 0.7$ exists for the very leading edge of waves of any frequency ($f_0$ to $f_8$), and also given that $\mathcal{T}_{WAF}$ either saturates at this value or decreases depending on the frequency, it indicates that when assuming the total energy budget on the planetary side of the TR to be enclosed in (a) a wave pulse or (b) a wave train, a greater proportion of the total wave energy will be transmitted through the TR for a wave pulse than for a wave train, especially at lower frequencies. Another noteworthy aspect is the impact of varying wave amplitude on transmittance. A preliminary analysis was also conducted by adjusting the wave amplitude at a specific frequency, revealing that such a variation resulted in negligible changes to the overall transmittance. Given the minimal influence observed, we determined that extending this line of analysis across all frequencies would yield limited insight. Consequently, we opted to focus our efforts on other aspects of the study as discussed in the following sections.

\subsection{Theoretical Estimates of the Transmittance\label{sec:leroy_theoretical_fit}}
An analytical description of the propagation of Alfv\'en waves in a radially stratified stellar atmosphere is not trivial. Various attempts have been made towards tackling this problem for non-WKBJ waves notable among which are \cite{Ferraro_plumpton_1958,Hollweg_1972, Hollweg_1986, Leroy_1979,Leroy_1980,Leroy_1981,Verdini_2005}.

Among the references mentioned, we will focus on the study by \cite{Leroy_1981} due to its ease of generalization. \cite{Leroy_1981} presents a systematic analytical method for calculating the transmittance of Alfvén waves in a stratified solar wind, which we will refer to as the `Leroy-1981' model henceforth. Although this methodology was originally envisioned to calculate the transmittance for waves propagating outward from the stellar boundary, the resulting transmittance is expected to be an intrinsic property of the medium itself and thus should be independent of the direction of wave propagation. Consequently, we evaluate its applicability to the Alfvén wave propagation considered in this study.

We begin by outlining some inherent assumptions of the model. The propagation medium is considered to consist of two isothermal layers connected by a temperature discontinuity. Specifically, the first isothermal layer, with a temperature $T_1$ extends from the stellar surface up to the base of the discontinuity at a height `h'. At this point, the temperature discontinuity increases the temperature from $T_1$ to a value of $T_2$. The second isothermal layer at a temperature $T_2$ then extends from this discontinuity up to a height `d'. The density profiles in the two isothermal layers are modeled as:
\begin{align}
    \rho_1 (z) &= \rho_{01} \rm{exp} \left[ \frac{-z}{H_1}\right] \qquad \rm{for} \qquad r_\star < z < h \\
    \rho_2 (z) &= \rho_{02} \rm{exp} \left[ \frac{-z}{H_2}\right] \qquad \rm{for} \qquad h < z < d \label{eq:exp2}
\end{align}
where the quantities $\rho_{01}$ and $\rho_{02}$ are the densities at the stellar boundary and the discontinuity respectively. 
The quantities $T_1$, $T_2$, $H_1$, $H_2$, `h' and `d' are then given as inputs to the model. The model then computes an analytical, frequency-dependent transmittance based on the methodology detailed in \cite{Leroy_1981}. 

 \begin{figure}
    \centering
    \includegraphics[width=1\columnwidth]{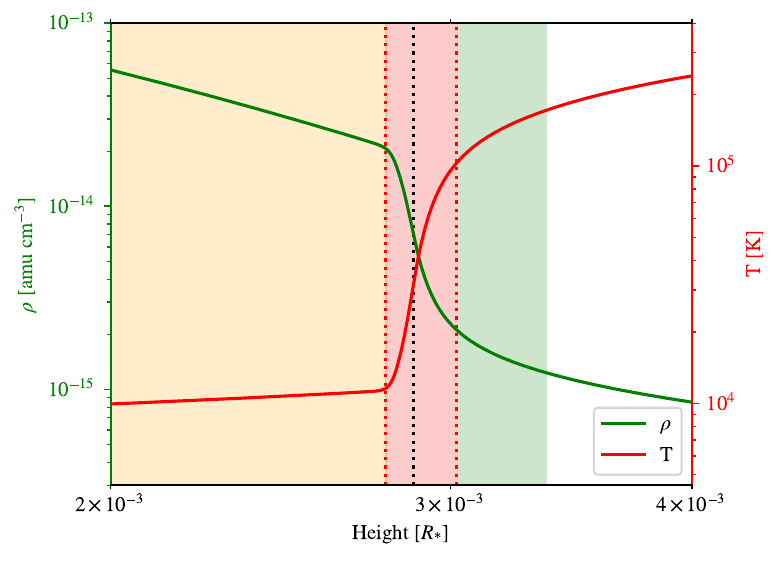}
    \caption{A highly magnified view near the transition region of the background solar wind profiles for $\rho$ and T. The yellow, red, and green shades represent three distinct regions, with parameters from each used as inputs to the Leroy-1981 model. For reference, we also include the black dotted line indicating the location of the TR, consistent with the previous plots.}
    \label{fig:leroy_regions}
\end{figure}

In order to apply the analytical model to the simulations presented in this paper, we progress as follows. First, similar to the analytical model, we divide up the numerical domain into three distinct regions. The first region extends from the stellar surface to the base of the TR denoted by the region shaded in yellow in figure \ref{fig:leroy_regions}. Note that the yellow region extends up to the stellar boundary on the left. The temperature profile in panel (a) of figure \ref{fig:background} clearly shows a sharp rise over a very short distance, which we broadly define as the TR. However, a detailed examination of the zoomed-in view of the same region in figure \ref{fig:leroy_regions}, reveals that the TR actually comprises two distinct sub-profiles: an initial steep rise followed by a more gradual increase. In the context of the Leroy-1981 model, we designate the steepest rise in the temperature profile as the `discontinuity'. This discontinuity is represented by the region shaded in red in figure \ref{fig:leroy_regions}. We now address the second exponential from the Leroy-1981 model within our numerical background model. We find that equation \ref{eq:exp2} produces a very poor fit of the region to the right of the discontinuity (see also, the appendix in \citealt{Reville_2018}). Therefore, for the sake of completeness, we select a very thin slice of this region, approximately 0.3 times the local density scale height, and treat it as a placeholder region analogous to the second exponential in the Leroy-1981 model. This region is denoted by the green slice in figure \ref{fig:leroy_regions}. Such an approach is also considered in \cite{Reville_2018} given the poor fit of the second exponential in their work as well. The dashed line profiles in Figure \ref{fig:WAF0.003} indicate minimal wave reflection as the waves approach the TR. It is therefore evident that the region to the right of the TR has negligible impact on wave reflection and the transmittance. Such a region is therefore redundant in the calculation of the $\mathcal{T}_{WAF}$ and its exclusion is therefore also deemed to be inconsequential. We also confirm that, upon applying the Leroy-1981 model algorithm to the background wind profile in this study, any change in the parameter `$H_2$', which results from fitting the second exponential, has a very negligible effect on the overall theoretical $\mathcal{T}_{WAF}$ profile.

Taking the above considerations into account, we fit our background solar wind profile to the assumptions of the Leroy-1981 model and obtain the parameters of the Leroy-1981 model as $\rm T_1 = 11365$ K, $\rm T_2 = 103232$ K, $\rm H_1 = 348$ km, $\rm H_2 = 6344$ km, h = 1921 km, d = 2326 km, $\rho_0 = 1.6\times 10^{-12}$ g cm$^{-3}$ and $\rm B_0 = 1.5$ G. We note that the values obtained here differ significantly from those used in the examples of \cite{Leroy_1981}. However, our parameters are dependent on the background wind model considered in this study and are therefore consistent with the wave propagation results presented. The analytical profile of $\mathcal{T}_{WAF}$ obtained from the above set of parameters is plotted as the blue solid line in figure \ref{fig:TC_leroy}. We see that the analytical profile does indeed closely follow the scatter points, i.e., values of the $\mathcal{T}_{WAF}$ obtained from the simulations in section \ref{sec:tc_sim}. The analytical profile also shows a strong attenuation of the $\mathcal{T}_{WAF}$ for the lower frequencies. The profile then shows an increase over a range of values, eventually attaining a saturation at an asymptotic value of $\mathcal{T}_{WAF} \sim 0.75$.

 \begin{figure}
    \centering
    \includegraphics[width=1\columnwidth]{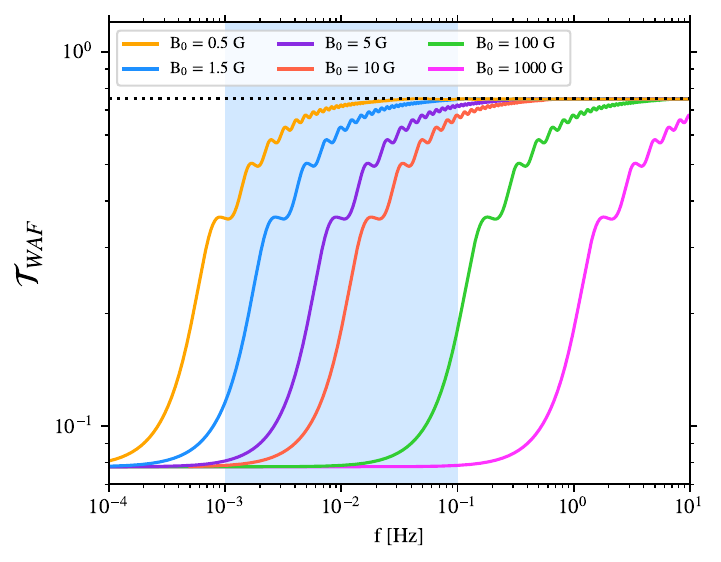}
    \caption{Analytical profiles of $\mathcal{T}_{WAF}$ obtained for different stellar magnetic field strengths. The blue profile corresponds to frequency range of the simulations explored in this study.}
    \label{fig:leroy_diffB}
\end{figure}

The excellent fit between the analytical profile and our simulation results opens a new avenue for exploration. It now gives us confidence that we can reasonably estimate $\mathcal{T}_{WAF}$ by simply varying the parameters of the analytical model, without fully depending on computationally expensive numerical simulations. Due to the numerous parameters needed to calculate the analytical model, varying all parameters simultaneously to analyze the behavior of the $\mathcal{T}_{WAF}$ profiles is extremely challenging. Therefore, we consider variations in specific individual parameters while keeping the others fixed, allowing us to assess how the $\mathcal{T}_{WAF}$ profiles depend on these individual parameters. 

We first vary the stellar magnetic field, considering values from slightly weaker than the Sun's magnetic field to strengths comparable to those of M dwarfs, which can reach kilo-Gauss levels. Figure \ref{fig:leroy_diffB} shows the analytical profiles of $\mathcal{T}_{WAF}$ obtained from the above exercise. The frequency range considered in this analytical exploration is significantly larger than the simulations performed in this study and the range considered in the numerical simulations is shaded in blue in figure \ref{fig:leroy_diffB} for a quick comparison. We choose to show a higher frequency range in for this plot to better capture and emphasize variations across a broader range of magnetic fields, which would not be as effectively represented with an abscissa range of [0.001-0.1] Hz.

It is seen that as the stellar magnetic field increases, lower frequencies tend to be reflected significantly, e.g., for a stellar magnetic field of 1 kG, all frequencies below $\sim$0.2 Hz have a $\mathcal{T}_{WAF}$ below 0.08 (8\%). The maximum $\mathcal{T}_{WAF}$ achievable by these profiles is independent of the stellar magnetic field strength and tends to be around 0.75. Notably, for higher stellar magnetic fields, only the frequencies towards the higher end of the spectrum are reflected the least.

Our next aim is to vary the surface temperature of the star and explore its effects on the analytical $\mathcal{T}_{WAF}$ profiles, however, this turns out to be slightly more involved. Although the model assumes an isothermal layer from the stellar surface to the discontinuity, the simulation reveals a temperature difference between these two points. Consequently, while the stellar surface temperature is $\sim$ 5977 K in the simulation, the temperature at the bottom of the discontinuity reaches around 11365 K. Using this data, we can make a crude approximation by considering a simple monotonous slope in temperature from the stellar surface to the discontinuity, allowing us to estimate the temperature at the discontinuity for various stellar surface temperatures which we can then use in the Leroy-1981 model. We consider four different spectral types of star, i.e., M, K G and F having typical temperatures of 3900 K, 5300 K, 6000 K and 7300 K respectively.

 \begin{figure}
    \centering    \includegraphics[width=1\columnwidth]{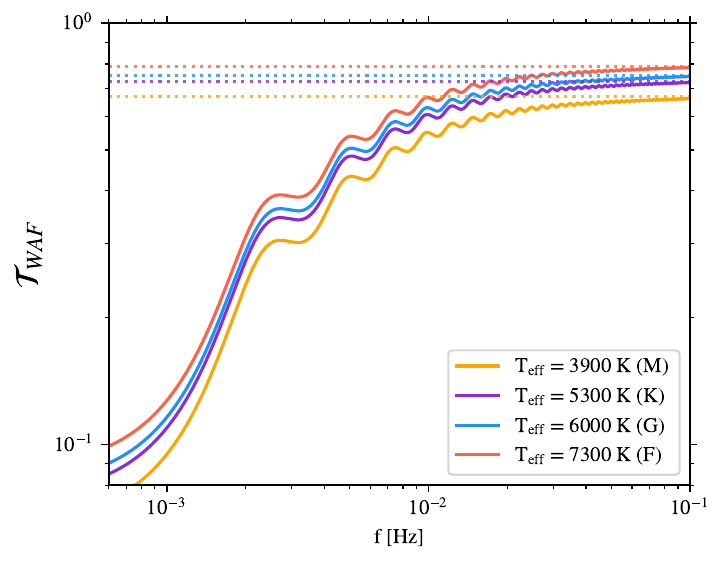}
    \caption{Analytical profiles of $\mathcal{T}_{WAF}$ obtained for different stellar surface temperatures. The letters in parentheses denotes the stellar spectral class representative of the temperature.}
    \label{fig:leroy_diffT}
\end{figure}
Figure \ref{fig:leroy_diffT} shows the analytical profiles of $\mathcal{T}_{WAF}$ obtained for different stellar surface temperatures. In the plots, the surface temperatures, denoted by $\rm T_{eff}$, serve as a proxy for the stellar spectral type. The corresponding spectral types are also indicated in the legend within parentheses. It is seen that the $\mathcal{T}_{WAF}$ profiles are relatively unaffected by changes in stellar surface temperature. However, the general trend shows that lower frequencies are transmitted slightly more as the stellar surface temperature increases. The asymptotic maximum of $\mathcal{T}_{WAF}$ also increases slightly with rising stellar surface temperature, reaching a value of approximately 0.78 for F-type stars. The  asymptotic maxima level of each spectral type is denoted by the correspondingly colored dashed line in figure \ref{fig:leroy_diffT}. As evident, this increase is minimal, however, this analysis relies on several assumptions and a thorough exploration of the entire parameter space is necessary for a precise assessment and will be achieved in a future work. Consequently, this part of the analysis should be considered as indicative of a general trend rather than a precise quantification. Such a trend may also not be directly applicable to stars with highly distinctive surface features, such as M-dwarfs, which can exhibit significant variations in temperature across their surfaces.

\subsection{Parametric decay of high frequency Alfvén waves\label{sec:PDI}}
We now explore another aspect of Alfvén waves that could potentially alter the overall transmittance of planetary waves. Parametric decay instability (PDI) is a well established phenomenon affecting Alfvén waves, where an incoming wave moving towards the star decays into an incoming compressive fluctuation and an outgoing (moving away from the star) Alfvén wave \citep{Shi_2017}. Alfvén waves are known to be susceptible to PDI when plasma $\beta \ll 1$ and are strongly stabilized for $\beta \sim 1$ and higher. It is also known that PDI and Alfv\'enic turbulence are interrelated. PDI generates large amplitude backscattered Alfv\'en waves, which interact with the parent waves and can lead to turbulence \citep{Tenerani_2017}. In addition to a dependence of the onset of PDI on the plasma-$\beta$ and the relative perturbation amplitude, \cite{Tenerani_2017} also highlighted that the high frequency waves are susceptible to PDI whereas their low frequency counterparts are stabilized.  Indeed the plasma beta within the region where the planetary waves are injected into the domain is of the order of $\sim 10^{-4}$ and therefore provides a suitable condition for the occurrence of PDI. While the primary focus of this study is not on PDI, it is essential to highlight its relevance here as the phenomenon plays a crucial role in assessing the transmittance for the high frequency end of the spectrum considered here. Specifically, we find that the highest frequency wave considered in the frequency set, i.e, f = 0.1 Hz exhibits signatures of PDI. \cite{Tenerani_2017} also reemphasized that PDI occurrence leads to fluctuations in the density and radial velocity profiles. To illustrate these fluctuations in our simulation domain, we plot the density and velocity perturbations occurring near the injection region, defined as:
\begin{align}
    \delta \rho &= \frac{\rho(t = 1.06 t_A) - \rho(0)}{\rho(0)}\label{eq:delrhoPDI}\\
    \delta v_r &= \frac{v_r (t = 1.06 t_A) - v_r(0)}{v_r(0)}\label{eq:delvPDI}
\end{align}

 \begin{figure}
    \centering    \includegraphics[width=1\columnwidth]{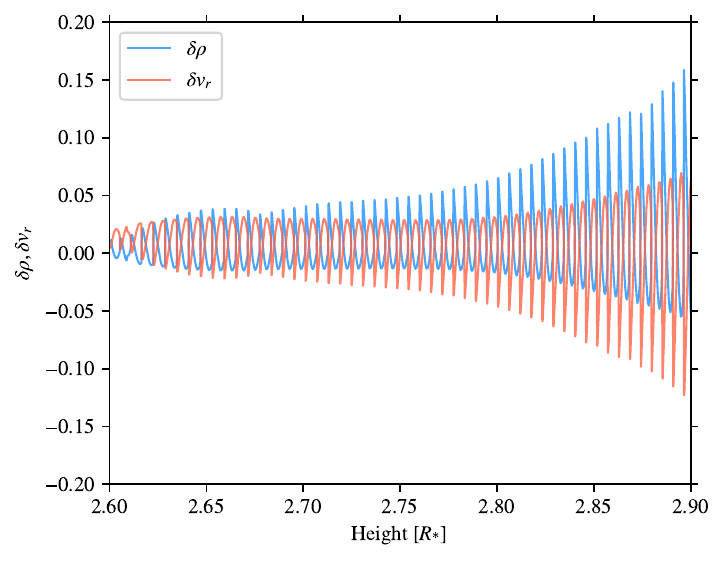}
    \caption{Density and velocity fluctuations obtained using equations \ref{eq:delrhoPDI} and \ref{eq:delvPDI} near the injection region for f = 0.1 Hz ($f_9$).}
    \label{fig:pdi_0.1}
\end{figure}

The presence of these perturbations is a clear indicator of PDI occurrence. Similar features have been observed in previous studies on PDI in Alfvén waves \citep{Tenerani_2017, Reville_2018}. Such signatures of PDI have been observed in two of the highest frequencies simulated in this study, namely, $f_8$= $5.99 \times 10^{-2}$ Hz and $f_9$ = 0.1 Hz. For the case of  $f_8$= $5.99 \times 10^{-2}$ Hz, PDI develops at a significantly later stage, possibly due to a much slower growth rate, thereby allowing us to calculate the $\mathcal{T}_{WAF}$ before the onset of PDI affects the WAF profile. For f= 0.1 Hz, however, the development of PDI occurs immediately after the waves are injected into the domain. As a result, the WAF profiles are significantly affected and the calculation of the $\mathcal{T}_{WAF}$ corresponding to the true frequency of the primary wave becomes impossible. The PDI converts the injected Alfvén waves into acoustic waves traveling towards the star due to the wind being just below supersonic at the location of the planetary orbit, as shown in panel (b) of Figure \ref{fig:background}. If the planetary orbit was in supersonic wind, the compressive waves would not propagate towards the star. Additionally, the process also generates backward-propagating Alfv\'en waves that move away from the star, carrying away a portion of the primary wave's energy in the process. In effect, the incoming WAF diminishes significantly by the time it travels from the location of the planet to the TR. 

 \begin{figure}
    \centering    \includegraphics[width=1\columnwidth]{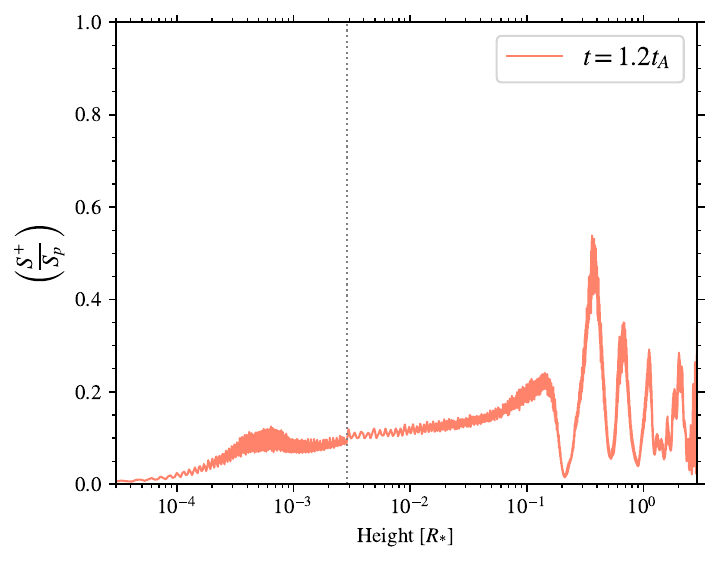}
    \caption{Normalized incoming ($S^+$) component of the WAF for f = 0.1 Hz ($f_9$) at t = 1.2 $t_A$, highlighting the effect of PDI on the WAF profile.}
    \label{fig:WAF_0.1}
\end{figure}

The red curve in figure \ref{fig:WAF_0.1} represents the instantaneous normalized incoming ($S^+$) WAF profile at t = 1.3 $t_A$. It is clear that the level of the instantaneous WAF reaching the stellar surface (left boundary of the plot) is extremely small and is practically zero. This is in contrast to the values obtained for higher frequencies in figure \ref{fig:TC_leroy} which sets an expectation of $\mathcal{T}_{WAF} \sim 0.75$ for such frequencies. Most of the WAF level drops very close to the injection region which is why we highlight the intense $\rho$ and $v_r$ fluctuations in such a zoomed-in region in figure \ref{fig:pdi_0.1}. Due to the non linear interactions between the backscattered Alfvén waves generated by PDI and the parent Alfvén waves, the overall WAF profile is also highly variable over time with small bursts of slightly higher $\mathcal{T}_{WAF}$ values. We therefore assert that in the presence of additional constraints in wave propagation, such as PDI, the resultant normalized WAF that finally reaches the stellar surface (also defined as the $\mathcal{T}_{WAF}$) drops significantly. This reduction is not due to reflection at the transition region but rather because PDI distributes most of the parent Alfvén wave's energy into secondary sonic waves (which are dissipative) and also into an Alfvén wave that propagates away from the star. The small amount of star-ward propagating WAF is therefore a result of the turbulent cascade between the counter-propagating daughter Alfv\'en waves and the primary monochromatic wave. This extremely low value of transmittance due to PDI is the reason why we omit the simulated $\mathcal{T}_{WAF}$ value for f= 0.1 Hz in figure \ref{fig:TC_leroy}.

\section{Implications for SPMI detectability\label{sec:observations_SPMI detect}}
Star-planet magnetic interactions occurring within the Alfvén surface of a star generate Alfvén waves that carry energy from the planetary obstacle toward the star. These Alfvén waves propagate along the magnetic field lines connecting the star to the planetary obstacle or its magnetosphere. The stellar transition region, characterized by a steep gradient in Alfv\'en speed, interacts with these waves, resulting in a partial transmission wherein a portion of the energy is reflected back. The most reliable indicators of SPMI to date are believed to be chromospheric hotspots generated by the energy carried by these Alfvén waves. It is therefore crucial to characterize the transmittance of these waves through the stellar transition region to determine the fraction of energy that successfully reaches the chromosphere. This study has focused on analyzing this interaction and has quantified this frequency-dependent transmission of Alfv\'en waves at the transition region. Such a frequency-dependent transmission has significant implications, particularly in observational contexts, which we will explore further below.

The frequency dependent transmission quantified in the results above leads us towards an intriguing inference: a frequency window exists, bounded by the obstacle size at the lower end and the frequency of onset of PDI at the upper end, within which Alfvén wave-mediated energy transfer is most efficient. The lower frequency limit is set by the physical constraint that an obstacle of a given size cannot generate waves below a certain threshold. While the classical upper frequency limit is technically defined by the ion-cyclotron frequency, the onset of PDI at a certain frequency in Alfv\'en waves prevents the parent waves from transferring energy effectively through the transition region beyond this frequency and upon the onset of PDI, the efficiency of energy transfer via Alfvén wave propagation drops significantly.
 
To better understand the total power budget available from star-planet magnetic interactions (SPMI) and the efficiency of Alfvén wave-mediated energy transfer, which ultimately determines the fraction of this power reaching the stellar chromosphere, we construct a grid of exoplanetary systems with varying stellar and planetary magnetic field strengths. The planetary magnetic fields ($\rm B_p$) are varied from 0.1 G to 500 G, while the stellar photospheric magnetic fields ($\rm B_{\star}$) range from 0.5 G to 1000 G within our grid. The lower limit of $\rm B_{\star}$ ensures that the planetary orbit remains within the star's Alfv\'en radius, and the upper limit reflects typical magnetic field strengths for M-dwarf stars. The range for $\rm B_p$ is based on the optimistic expected magnetic field strengths of exoplanets \citep{Yadav_2017}, including that of hot-Jupiters. As there are a large number of parameters that can influence SPMI, it is imperative that we keep some of them fixed in order to simplify the rest of the calculations. As such, in this study, we consider that the stars with varying magnetic fields have approximately similar atmospheric temperature and density profiles as that of the Sun. We defer a proper self-consistent characterization of these to a future work. We also consider for now that the planetary orbital radius is fixed at 4 $\rm R_{\star}$. Changing the orbital radius can have two implications, (a) it can dictate whether the planet is within or outside the Alfv\'en surface thereby switching on/off any Alfv\'en waves propagating towards the star and (b) it would change the orbital speed thereby influencing the lower limit of the frequency that can be produced by that obstacle. A similar analysis to that presented in the following paragraphs can be readily applied to exoplanets with different orbital radii when investigating specific exoplanetary systems motivated by observations.

We begin by estimating the efficiency of power transfer, following the outlined analysis. For each point on the [$\rm B_p$, $\rm B_\star$] grid, we calculate the effective obstacle size of a magnetized planet within the sub-Alfv\'enic stellar wind using the magnetic pressure balance equation:
\begin{equation}\label{eq:eff_obs_size}
   \rm R_{obs} = R_p  \left(\frac{B_p}{B_{\star (r=r_{orb})}}\right)^{\frac{1}{3}}
\end{equation}
where $\rm R_p$ represents the planetary radius, which is assumed to be equivalent to one Jupiter radius in this specific case. The planetary radius represents the minimum obstacle size when the planetary magnetic field approaches zero. $\rm B_p$ represents the surface magnetic field of the planet at the equator and $\rm B_{\star (r=r_{orb})}$ represents the magnitude of the stellar magnetic field at the location of the planet. $\rm B_{\star (r=r_{orb})}$ is derived from the stellar photospheric magnetic field considering that the field strength falls off with an inverse square dependence with distance. The obstacle size determines the lower limit of the frequency ($f_{low}$) of the Alfvén waves that the planet can possibly generate and is calculated using the expression:
\begin{equation}
    \rm f_{low} = \frac{v_{orb}}{2  R_{obs}} \qquad [Hz]
\end{equation}
where $\rm v_{orb}$ represents the orbital speed of the planet and is dependent on the stellar mass and the radius of the orbit. The upper frequency bound for the Alfv\'en waves is considered to be demarcated by the local ion cyclotron frequency ($f_{high } = \Omega_{ci}$). We now turn our attention to existing estimates of the total power budget generated by SPMI. An analytical model proposed by \citet{Saur_2013} (and reproduced satisfactorily in numerical simulation by \citealt{Strugarek_2016}) provides an expression for this power, which is given by: 
\begin{equation}
    \rm S_{SPMI} = 2\pi R_{obs}^2 v_A \frac{(\alpha M_A B_{\star} (r=r_{orb}) \cos \theta)^2}{\mu_0} \rm \qquad [watts]
\end{equation}
where $\rm v_A$, $\rm M_A$, $\rm B_{\star (r=r_{orb})}$ represent the Alfv\'en speed, Alfv\'enic Mach number, stellar magnetic field at the orbital location of the exoplanet, respectively. The parameter $\alpha$ quantifies the interaction efficiency, indicating how much of the Poynting flux at the planet’s location is directed back toward the star via the Alfvén wings, whereas, the quantity $\theta$ represents the relative alignment between the stellar and planetary magnetic fields For the purposes of this analysis, both $\alpha$ and $\cos\theta$ are set to unity to estimate the maximum possible power output. For each point on the [$\rm B_p$, $\rm B_\star$] grid, we distribute this available power into frequency bins bounded at the lower end by $f_{low}$ and at the upper end by $f_{high}$. For simplicity, we consider a Kolmogorov type power-law spectrum of the power throughout this frequency range  and the division of the total power ($\rm S_{SPMI}$) into the frequency bins is done on the basis of the following expression:
\begin{equation}
    S_{\rm SPMI} = \int_{f_{low}}^{f_{high}} P(f)\, df = \int_{f_{low}}^{f_{high}} \mathcal{K} \, f^{\frac{-5}{3}} \, df
\end{equation}
where the constant $\mathcal{K}$ can be calculated from the analytical integration of the integrand. Next, since the Leroy-1981 model provided an excellent fit of the simulation data as demonstrated in section \ref{sec:leroy_theoretical_fit}, we calculate the transmittance for each frequency bin $\mathcal{T}_{WAF} (f)$ from the previous step from this model. Using this transmittance, we then estimate the power transmitted by each of these frequency bins  as an element-wise product $P_{tr} (f) = \mathcal{T}_{WAF} (f) \otimes P(f)$. The quantity $ P_{tr} (f)$ is then summed up over the frequency bins to finally determine the total power that is actually transmitted to the stellar chromosphere through the transition region ($\rm{S_{tr}} = \mathit{\sum P_{tr} (f)}$). For a particular stellar and planetary magnetic field strength, the efficiency of this transmission is then quantified using the following expression:
\begin{equation}
    \rm S_{\%} = \frac{S_{tr}}{S_{SPMI}} \times 100
\end{equation}
While calculating the transmittance, we also keep in mind that certain frequencies can be unstable to PDI. It is important to note that the occurrence of PDI in high-frequency Alfv\'en waves depends on thresholds for plasma $\beta$ and the parent Alfv\'en wave amplitude and frequency \citep{Li_2022, Reville_2018}. In the present work, the wave amplitudes has been fixed as mentioned in section \ref{sec:numer_set}. The growth rate of PDI can be given as $\gamma_{max} = [\omega_0 (a(1-\sqrt{\beta})^{1/2})]/[2\beta^{1/4}(1+\sqrt{\beta})]$ wherein $a$ is the relative perturbation amplitude  and $\omega_0$  is the wave frequency \citep{Jayanti_1993, Reville_2018}. It can be seen that within the low-$\beta$ regime, the expression reduces to $\gamma_{max}\propto \omega_0 \beta^{-1/4}$. For each point on the [$\rm B_{p}$, $\rm B_{\star}$] grid, we calculate a threshold frequency for PDI onset using the following logic. In the simulated system, with a certain fixed value of $\beta$, PDI occurs rapidly at f$\sim$ 0.1 Hz. This information essentially reveals a value of $\gamma_{max}$ that would definitively give rise to a rapid PDI onset. For systems with different $\beta$ values, we therefore leverage this value of $\gamma_{max}$ to obtain a threshold frequency ($\omega_0$) of PDI onset by plugging it into the expression $\gamma_{max}\propto \omega_0 \beta^{-1/4}$. In the frequency distribution of the total Poynting flux budget, all frequencies above this threshold for each point in the [$\rm B_p$, $\rm B_{\star}$] grid are assigned a transmittance value of $\mathcal{T}_{WAF} \sim$ 0.002, inferring from section \ref{sec:PDI}. Since the effective obstacle size, given by equation \ref{eq:eff_obs_size} depends on the planetary radius as a lower bound, the resulting efficiency profile for the [$\rm B_p$, $\rm B_\star$] grid also inherently possesses this dependence. Therefore, we present the efficiency profiles within the [$\rm B_p$, $\rm B_\star$] grid for three different planetary sizes; $\rm R_p$ = 0.1 $\rm R_E$, representing a small planet; $\rm R_p$ = 1 $\rm R_E$, representing a moderately sized planet; and $\rm R_p$ = 1 $\rm R_J$, representing a large planet comparable to Jupiter.

 \begin{figure*}
    \centering    \includegraphics[width=1.0\linewidth]{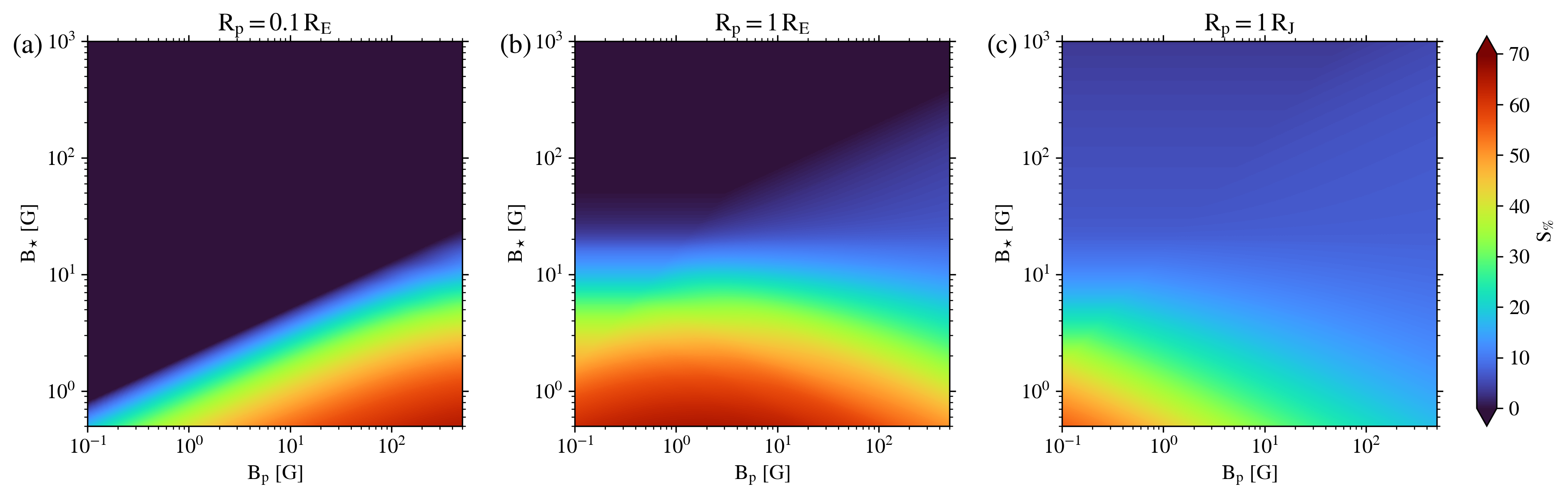}
    \caption{The efficiency of power transmission to the stellar chromosphere within a parameter space of [$\rm B_p$, $\rm B_\star$] for different exoplanetary sizes.}
    \label{fig:S_percent}
\end{figure*}
It is immediately apparent that there is a specific region within the entire [$\rm B_p$, $\rm B_\star$] grid where energy transfer is most efficient. For small planetary sizes, $\rm \sim0.1 R_E$, the most efficient power transfer occurs at higher planetary magnetic field strengths. For intermediate-sized planets comparable to the size of the Earth, efficiency peaks at intermediate magnetic field strengths. Conversely, for large planets comparable to the size of Jupiter, the most efficient power transfer is observed at lower planetary magnetic field strengths. The above analysis leads to an inference that there in fact exists an optimal obstacle size that corresponds to small planets with large magnetospheres or large planets with small magnetospheres which eventually gives rise to the most optimal transmission of Alfvén waves through the transition region. In panels (a) and (b) of figure \ref{fig:S_percent}, there are regions where the efficiency approaches zero. These areas arise because, for certain combinations of stellar and planetary magnetic field values, the lowest frequency ($f_{low}$) that the obstacle can generate is sufficiently high to make the waves unstable due to PDI. As a result, the overall efficiency of power transmission through the entire spectrum of these waves is effectively zero. In contrast, for larger planetary sizes, the lowest frequency that the obstacle can generate is lower than the frequency required to trigger PDI, which is why regions of near-zero efficiency are absent in panel (c). Additionally, it is important to note that the maximum efficiency reaches approximately 70\% for specific combinations of planetary sizes, planetary magnetic fields, and stellar magnetic fields. However, for most of the [$\rm B_p$, $\rm B_\star$] parameter space considered, the efficiency is relatively lower.

\begin{figure}
    \centering    \includegraphics[width=1\columnwidth]{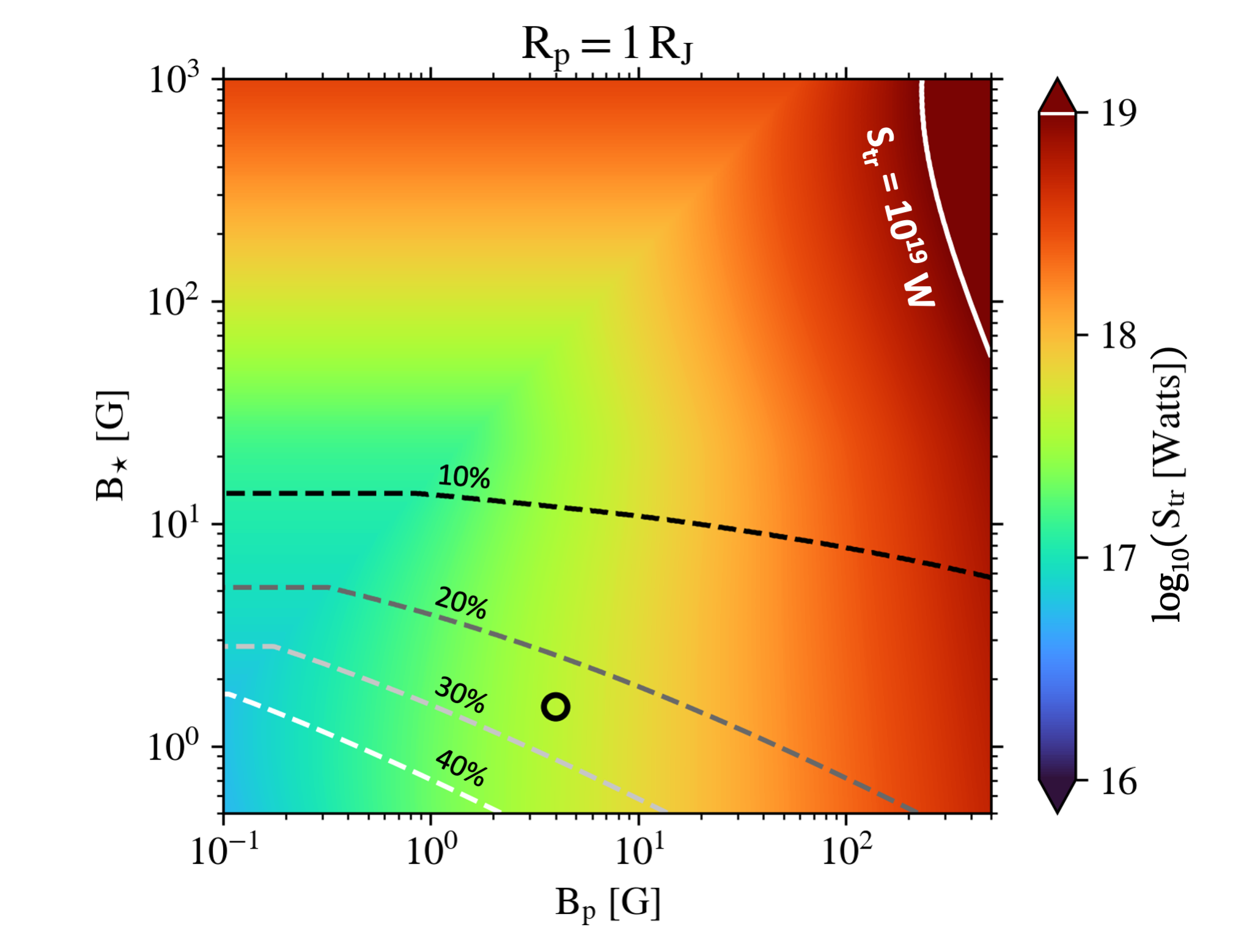}
    \caption{Power transmitted to the chromosphere, as estimated using the transmission efficiencies calculated from this work along with the total power budget from the \citet{Saur_2013} model. The dashed lines indicate the contours of transmission efficiency ($\rm S_\%$), while the solid contour delineates a transmitted power level of $10^{19}$ watts.}
    \label{fig:S_tot_saur}
\end{figure}

We now focus on quantifying the amount of power transmitted to the stellar chromosphere, using the power estimates provided by the \citet{Saur_2013} model and the efficiencies determined in this study. For a Jupiter sized planet at an orbital radius of 4 $R_{\star}$, figure \ref{fig:S_tot_saur} represents the power transmitted to the chromosphere as a function of the planetary and stellar magnetic field strengths. 
The dashed lines in the figure \ref{fig:S_tot_saur} represent transmission efficiency contours having values of 10\%, 20\%, 30\% and 40\% wherein lighter shades represent higher values. The black circle represents the position of a planet with an equatorial magnetic field of 4 G (Jupiter like) orbiting around a star having an average surface magnetic field of 1.5 G (Sun like). The total power transmitted to the chromosphere for such a planet is calculated to be $4 \times 10^{17}$ W. And the efficiency of power transmission at this location in the [$\rm B_p$, $\rm B_\star$] grid is estimated at 24\%. 

As noted by \citet{Lanza_2013} and \citet{Cauley_2019}, the total power generated by SPMI, as estimated by \citet{Saur_2013}, is insufficient to account for the powers observed in chromospheric hotspots, which exhibit levels on the order of $10^{20}$ to $10^{21}$ W \citep{Shkolnik_2005, Lanza_2013, Cauley_2019}. The white contour in the upper right of the plot in figure \ref{fig:S_tot_saur} represents a power level of $10^{19}$ W. It is evident that even achieving power levels on the order of $10^{19}$ W at the chromosphere necessitates a combination of extremely high stellar and planetary magnetic fields. We therefore turn our attention to an alternate model of SPMI proposed by \citet{Lanza_2013} wherein the total power generated by such interaction is given by:
\begin{equation}\label{eq:Lanza_power}
   \rm  S_{tot} = \frac{2 \pi f_{AP} R_p^2 B_p^2 v_{rel}}{\mu_0} \rm \qquad [watts]
\end{equation}
where $\rm  R_p, B_p$ and $\rm  v_{rel}$ represent the exoplanetary radius, the exoplanetary polar magnetic field and the relative velocity between the exoplanet and the stellar wind, respectively. As a first approximation, we set $\rm  v_{rel}$ to be equal to the orbital speed of the planet for now. This will be set to be the vector addition of the orbital speed and the wind speed in the following sections when observations of exoplanets will be considered. The quantity $f_{AP}$ represents the fraction of the planetary surface magnetically connected to the stellar field and is given by:
\begin{equation}
    \rm f_{AP} = 1- \left( 1-\frac{3 \zeta^\frac{1}{3}}{2+\zeta}\right)^{\frac{1}{2}}
\end{equation}
where $\zeta$ is defined as:
\begin{equation}
    \rm \zeta = \frac{B_{\star (r= r_{orb})}}{B_p}
\end{equation}
We highlight here that the efficiencies shown in figure \ref{fig:S_percent} are independent of the magnitude of the available power budget, and therefore, the efficiency plots are equally valid for the power budgets calculated by the \citet{Saur_2013} and the \citet{Lanza_2013} models. Having said that, we move on to show the power transmitted to the chromosphere within our [$\rm B_p$, $\rm B_\star$] grid when the total available power budget from SPMI is calculated using the \citet{Lanza_2013} model for a Jupiter sized magnetized planet at an orbital radius of 4 $\rm R_{\star}$.
 \begin{figure}
    \centering    \includegraphics[width=1\columnwidth]{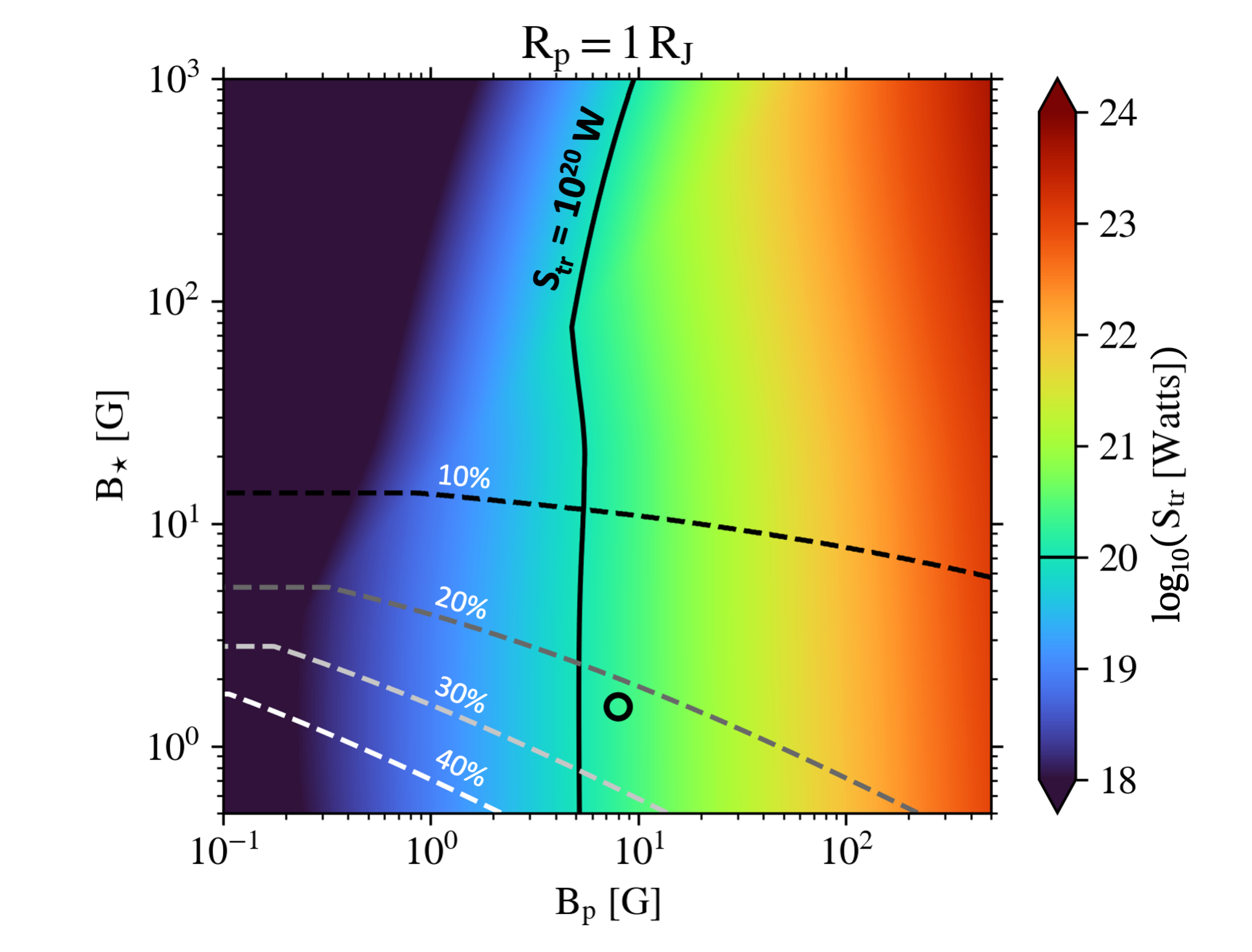}
    \caption{Power transmitted to the chromosphere, as estimated using the transmission efficiencies calculated from this work along with the total power budget from the \citet{Lanza_2013} model. The dashed lines indicate the contours of transmission efficiency ($\rm S_\%$), while the solid contour delineates a transmitted power level of $10^{20}$ watts in this case.}
    \label{fig:S_tot_lanza}
\end{figure}
As expected, the magnitudes of the total power generated by SPMI, estimated using the \citet{Lanza_2013} model is significantly higher than the values obtained from the \citet{Saur_2013} model. Similar contours of 10\%, 20\%, 30\% and 40\% efficiencies are shown in this figure as well. The solid black contour represents a level of $10^{20}$ W. The black circle once again represents the position of a planet with a polar magnetic field of 8 G (Jupiter like) orbiting around a Sun like star having an average surface magnetic field of 1.5 G. The total power transmitted to the chromosphere in this case is calculated to be $1.9 \times 10^{20}$ W which aligns more closely with the observed powers of chromospheric hotspots reported by \citet{Shkolnik_2005} and \citet{Cauley_2019}. It is evident that, despite the efficiency factor associated with power transmission through the stellar transition region, the power reaching the chromosphere is sufficient to account for the tentative observations of SPMI power in chromospheric hotspots. The efficiency of power transmission at this location in the [$\rm B_p$, $\rm B_\star$] grid is estimated at 22\% and this slight change in efficiency is due to the position of the point on the [$\rm B_p$, $\rm B_\star$] wherein, the \citet{Saur_2013} model considers an equatorial magnetic field strength of the planet whereas the \citet{Lanza_2013} model considers a polar magnetic field strength \citep{Cauley_2019}. 

 \begin{figure*}
    \centering    \includegraphics[width=1.0\linewidth]{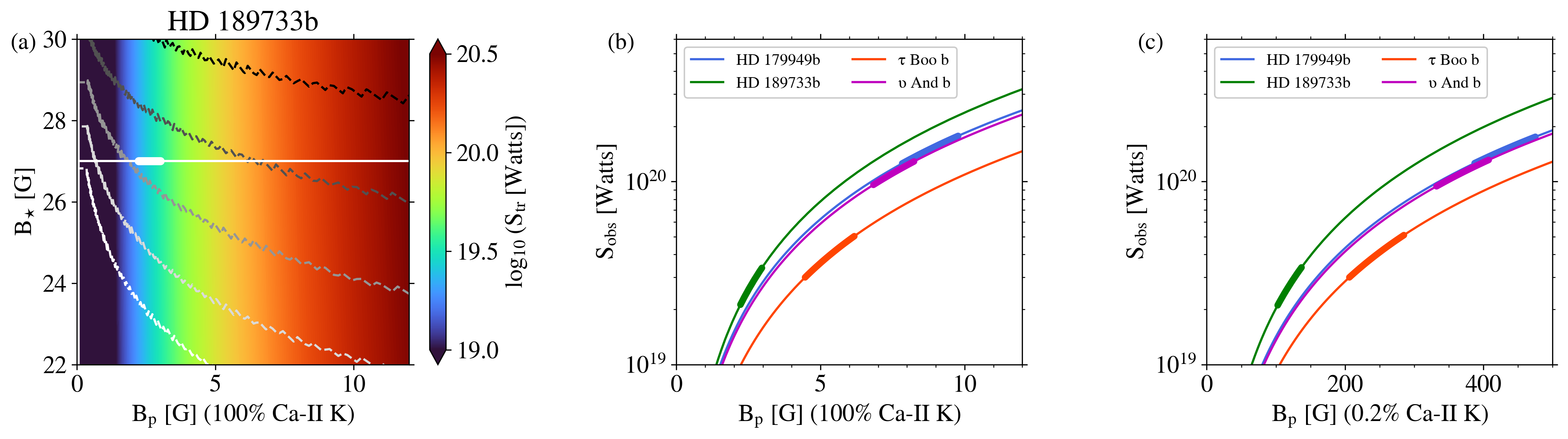}
    \caption{Panel (a) shows the variation of the transmitted Poynting flux for HD 189733b on a [$\rm B_\star$, $\rm B_p$] grid. The dashed contour lines show transmission efficiencies of (from the top) 8.0\%, 8.1\%, 8.2\%, 8.3\% and 8.4\% with lighter shades representing higher values. Panels (b) and (c) illustrate the dependence of the required planetary magnetic field strengths to account for the observed power emitted from certain exoplanetary systems suspected of exhibiting SPMI. Panel (b) presents a lower limit for the magnetic field strength, assuming that all the energy reaching the chromosphere is emitted in the Ca-II K band. In contrast, panel (c) offers a more realistic scenario, where the fields are calculated considering the constraint that 0.2\% of the energy reaching the chromosphere is emitted in the Ca-II K band. }
    \label{fig:observations}
\end{figure*}

The plots presented so far can offer valuable insights for target selection to observe potential star-planet magnetic interactions. Given the vast diversity within the exoplanetary population, it is essential to identify systems most likely to exhibit such interactions. This selection process should focus on two key factors: the efficiency of energy transfer between the planet and the star, and the total power transmitted to the stellar surface via Alfvén waves, which constrains the power available for emission as SPMI signatures. A degree of ambiguity remains in the latter case, as evidenced by the total transmitted power calculated using the models of \citet{Saur_2013} and \citet{Lanza_2013}. Although the \citet{Saur_2013} model has been verified through numerical simulations and accounts well for the power observed in planet-satellite interactions in the solar system, it produces at least an order of magnitude less power than required to match the observations. On the other hand, the \citet{Lanza_2013} model can account for the observed power, but it has yet to be validated by numerical simulations. In any case, based on the efficiencies of transmission quantified so far, which are independent of the model used to calculate the SPMI power budget, one can identify exoplanetary systems that are optimal for detecting magnetic interaction signatures and an informed selection of such systems with favorable conditions in terms of both energy transfer efficiency and magnitude of available power budget would enhance the likelihood of successful SPMI detections.

We now turn our attention to several observational cases where SPMI has been tentatively detected, and delve into the insights that can be drawn from the analysis conducted thus far. It's important to note that in the figures \ref{fig:S_percent}, \ref{fig:S_tot_saur} and \ref{fig:S_tot_lanza}, the planetary orbit was kept fixed at $\rm 4 R_{\star}$. To accurately construct comparable plots for various known exoplanets, we must re-calculate the [$\rm B_p$, $\rm B_{\star}$] variation of the transmitted power for different planetary orbital radii. We performed the following additional steps for that. \citet{Cauley_2019} recently reported tentative detections and estimates of SPMI power for four known exoplanetary candidates. We begin by focusing on these four candidates, with the details used for our calculations provided in table \ref{tab:exoplanets}

\begin{table*}
\caption{\begin{tiny} Stellar and planetary parameters of some well known exoplanetary systems where SPMI has tentatively been observed.\end{tiny} }
    \label{tab:exoplanets}
\centering
\begin{tiny}
\begin{tabular}{ ||c|c|c|c|c|c|c|c|| }
\hline
Exoplanet name  & $\rm R_\star$ [R$_{\odot}$]& $\rm P_\star$ [days] &$\rm B_\star$ [G] & $\rm R_{p}$ [$\rm R_J$] & $\rm P_{orb}$ [days] & $\rm a_{orb}$[AU] &  $\rm S_{Ca-II K}$ [$\times 10^{20}$ watts]\\ \hline
HD179949b & 1.23 & 11.0 & 3.2 & 1.22 & 3.0925 & 0.044 &$1.53\pm0.27$ \\
HD189733b & 0.76 & 11.9 & 27.0 & 1.14 & 2.2186 & 0.031 &$0.28\pm0.07$ \\
$\tau$ Boo b & 1.48 & 3.7 & 2.6 & 1.13 & 3.3124 & 0.048 &$0.41\pm0.11$\\
$\upsilon$ And b & 1.64 & 12.0 & 2.5 & 1.25 & 4.6170 & 0.059 & $1.14\pm0.19$\\
\hline
\end{tabular}
\end{tiny}
\tablefoot{\begin{tiny} $\rm R_\star$, $\rm P_\star$, $\rm B_\star$, $\rm R_{pl}$,  $\rm P_{orb}$ and $\rm a_{orb}$ represent the stellar radius, stellar rotation period, stellar magnetic field, the radius of the exoplanet, the orbital period of the exoplanet and the semi major axis of this orbit respectively. $\rm S_{Ca-II K}$ represents the SPMI power observed in Ca-II K band. The values are as reported in \citet{Cauley_2019} and have been used as input parameters in this work.\end{tiny}}
\end{table*}

The first significant variation from the previous analyses to these exoplanets is their orbital radius, which differs for each planet. For simplicity, and knowing that the orbits have very low eccentricities, we consider circular orbits with a radius equal to the semi-major axis ($\rm a_{orb}$) provided in table \ref{tab:exoplanets}. We also assume that at the planet's location, the stellar wind's azimuthal speed is nearly equal to the star's rotational speed. Therefore, the relative rotational velocity in the orbital direction is the difference between the Keplarian velocity of the planet and the star's rotational velocity at the orbit's location if one considers a corotating stellar atmosphere. The quantity $\rm v_{rel}$ in equation \ref{eq:Lanza_power} is therefore taken to be the vector sum of this relative azimuthal velocity and the radial velocity of the stellar wind at the location of the planet. We focus exclusively on the \citet{Lanza_2013} model for flux quantification, as it is currently the only model capable of accounting for the magnitude of observed chromospheric hotspot fluxes, despite its validity not yet being supported by numerical simulations. Under the above assumptions, we construct plots similar to figure \ref{fig:S_tot_lanza} for the four exoplanets highlighted in table \ref{tab:exoplanets}. One such plot is shown for the exoplanet HD189733b in panel (a) of figure \ref{fig:observations}. We draw attention to the fact here that panel (a) only shows a very small portion of the [$\rm B_p$, $\rm B_\star$] space considered in the previous analyses with linear axes scaling, however, the color of the transmitted power remains in the logarithmic scale. The dashed contours in panel (a) represents the efficiency of transmission within this parameter space. The five contour lines shows transmission efficiencies of 8.0\%, 8.1\%, 8.2\%, 8.3\% and 8.4\% starting from the top with lighter shades representing higher values. As anticipated from the trends observed in figure \ref{fig:leroy_diffB}, the efficiency decreases as the stellar magnetic field strength increases. The efficiencies also decrease with increasing planetary magnetic fields. This can be attributed to higher planetary magnetic fields leading to larger obstacle sizes or lower minimum wave frequencies. As observed in figure \ref{fig:TC_leroy}, transmittance decreases with a reduction in wave frequency, leading to this trend. Overall, for the planetary parameters of HD189733b, the predicted transmission efficiency is $\sim$ 8\%.

Our objective here is to attempt to provide constraints on the planet's magnetic field, similar to the approach taken by \citet{Cauley_2019}, but with the added consideration of the efficiency factor of power transmission through the transition region of the star. This inclusion would provide new lower limits for the planet's magnetic field. With appropriate constraints on the stellar average surface magnetic field strengths, the 2D plot shown in panel (a) of figure \ref{fig:observations} can, in principle, be condensed into a 1D plot, illustrating the variation of the transmitted Poynting flux as a function of the planetary magnetic field for a specific value of the stellar surface magnetic field strength.The white horizontal line in panel (a) of figure \ref{fig:observations} indicates the level of stellar magnetic field strength for HD189733b. While observations of stellar magnetic field strength do carry some uncertainties, for this particular analysis, the uncertainties are ignored because as can be seen in panel (a), shifting the horizontal line slightly vertically would not significantly alter the variation in transmitted Poynting flux profile along the line. We do, however, account for the uncertainties in the observed value of the emitted power within the chromospheric hotspots. In panel (a) of figure \ref{fig:observations}, the thick segment of the horizontal line represents the range of observed power values as indicated by the background color profile.We now examine the power profile along horizontal lines corresponding to the stellar magnetic field levels of all the exoplanets listed in table \ref{tab:exoplanets} and plot the variation in transmitted power with the planetary magnetic field strength for these planets. This dependence is shown in panels (b) and (c) of figure \ref{fig:observations}. Each line is colored according to the legends in panels (b) and (c), and the thicker portions of the curves represent the range of observed stellar hotspot power in the Ca-II K band for different considerations (elaborated in the following paragraphs). We emphasize that the powers highlighted on the ordinate represent the transmitted powers to the stellar chromosphere, taking into account the efficiency of transmission through the stellar transition region. For the observed power range, it is indeed possible to have constraints on the planetary magnetic field. We first consider a scenario where all of the power reaching the stellar chromosphere is emitted in the Ca-II K band. While this is not physically accurate, since only a portion of the power reaching the chromosphere is emitted in this band, such an approach is used to establish lower limits for the planetary magnetic field strengths \citep{Cauley_2019}. 
\begin{table*}[]
\caption{\begin{tiny}A comparison of the results obtained in \citet{Cauley_2019} and the results obtained in this study.\end{tiny}}
\centering
\begin{tiny}
\begin{tabular}{|c|cc|cccc|}
\hline
\multirow{2}{*}{Exoplanet name} & \multicolumn{2}{c|}{\citet{Cauley_2019}}                                             & \multicolumn{4}{c|}{This work}                                                                                                                                                         \\ \cline{2-7} 
                                & \multicolumn{1}{c|}{$\rm B_{p (100\% Ca-II K)}$} & $\rm B_{p (0.2\% Ca-II K)}$ & \multicolumn{1}{c|}{$\rm B_{p (100\% Ca-II K)}$} & \multicolumn{1}{c|}{$\rm S_{\% (100\% Ca-II K)}$} & \multicolumn{1}{c|}{$\rm B_{p (0.2\% Ca-II K)}$} & $\rm S_{\% (0.2\% Ca-II K)}$ \\ \hline
HD179949b                       & \multicolumn{1}{c|}{$1.9\pm 0.7$}                & $86\pm29$                   & \multicolumn{1}{c|}{$8.8\pm1.0$}                 & \multicolumn{1}{c|}{11.9}                         & \multicolumn{1}{c|}{$435.3\pm47.9$}              & 9.5                          \\
HD189733b                       & \multicolumn{1}{c|}{$0.4\pm0.1$}                 & $20\pm7$                    & \multicolumn{1}{c|}{$2.6\pm0.4$}                 & \multicolumn{1}{c|}{8.2}                          & \multicolumn{1}{c|}{$120.9\pm18.4$}              & 7.9                          \\
$\tau$ Boo b                    & \multicolumn{1}{c|}{$2.7\pm0.9$}                 & $117\pm38$                  & \multicolumn{1}{c|}{$5.4\pm0.9$}                 & \multicolumn{1}{c|}{9.4}                          & \multicolumn{1}{c|}{$248.0\pm41.8$}              & 8.5                          \\
$\upsilon$ And b                & \multicolumn{1}{c|}{$1.9\pm1.8$}                 & $83\pm77$                   & \multicolumn{1}{c|}{$7.6\pm0.8$}                 & \multicolumn{1}{c|}{11.8}                         & \multicolumn{1}{c|}{$373.4\pm41.2$}              & 9.5                          \\ \hline
\end{tabular}
\label{tab:Bexo_observaitons}
\end{tiny}
\end{table*}

The first column in table \ref{tab:Bexo_observaitons} lists the names of the exoplanets. The second and third columns present the planetary magnetic field strengths as determined by \citet{Cauley_2019}. The second column titled `$\rm B_{p (100\% Ca-II K)}$' shows the lower limit of the magnetic field strength, assuming 100\% of the emitted energy is in the Ca-II K band. The third column named `$\rm B_{p (0.2\% Ca-II K)}$' offers a more realistic estimate, where the magnetic field strength is calculated based on only 0.2\% of the available power being emitted in the Ca-II K band. As shown in these columns derived from \citet{Cauley_2019}, the absolute lower limit of the planetary magnetic fields required to explain the emitted power ranges from 0.4 G to 1.9 G for the candidates considered here, while $\rm B_{p (0.2\% Ca-II K)}$ falls between 20 G and 117 G. The $\rm B_{p (0.2\% Ca-II K)}$ values obtained by \citet{Cauley_2019} are consistent with the expected range for hot Jupiters as highlighted by \citet{Yadav_2017}. However, in the scenario where the power from SPMI are carried to the star by Alfvén waves, these values do not take into account the reflection of Alfvén waves at the stellar transition region and the associated efficiency factor. To address this, we now consider the efficiency of energy transfer through the transition region which is a critical step for Alfvén waves to propagate to the stellar chromosphere, and recalculate the planetary magnetic field strengths. The fourth column of table \ref{tab:Bexo_observaitons} similarly presents our obtained $\rm B_{p (100\% Ca-II K)}$ values as a lower limit for the planetary magnetic field strength, factoring in the efficiency of power transmission at the transition region. The constraints on planetary magnetic field strengths for this scenario is shown in panel (b) of figure \ref{fig:observations}. The vertical range of the thick segments of each line corresponds to the observed chromospheric power values reported in \citet{Cauley_2019}, while the horizontal extents represent the required planetary magnetic field strengths to generate these power levels. The obtained values now range from 2.6 G to 8.8 G for the sample set, which, although higher, remain comparable to the results of \citet{Cauley_2019}. However, when calculating the magnetic field strengths based on the more realistic $\rm B_{p (0.2\% Ca-II K)}$ metric, the required values to account for the observed power are significantly higher, ranging from 121 G to 435 G. 
This is similarly plotted in panel (c) of Figure \ref{fig:observations}, which mirrors panel (b) but accounts for the assumption that 0.2\% of the total power is emitted in the Ca-II K band. Indeed \citet{Yadav_2017} had highlighted that such high values are possible for very massive hot-Jupiters  with masses $\sim 10 M_J$, however, that is not the case for the candidates considered here. Additionally, as elaborated before, our approach also allows us to determine the efficiency of energy transfer through the transition region for these systems. This efficiency depends on the exoplanet's position within the [$\rm B_p$, $\rm B_\star$] parameter space and is detailed in the columns titled $\rm S_{\% (100\% Ca-II K)}$ (for the lower limit of the planetary magnetic field strengths) and $\rm S_{\% (0.2\% Ca-II K)}$ (for the realistic estimate of 0.2\% of the SPMI power being emitted in the Ca-II K band), corresponding to each of the magnetic field strengths calculated in this study. As indicated by these columns, the efficiency for all the candidates examined in this study is $\sim$10\% which means that approximately 90\% of the total energy carried by the Alfv\'en waves will be reflected back and will never actually reach the chromosphere in order to generate emissions. Therefore, even when calculating the SPMI power using the \citet{Lanza_2013} model, accounting for efficiencies at the transition region still yields unrealistic values for the planetary magnetic field strengths. The need for unrealistically high planetary magnetic field strengths to account for the observed power when using the \citet{Saur_2013} model was a key factor in preferring the \citet{Lanza_2013} model. We now find ourselves confronted with a similar issue again. 

Such discrepancies indeed highlight the need to move beyond simplistic estimations of the power budget generated by SPMI. Even within the models considered, a large planetary magnetosphere would result in stronger stellar magnetic field strengths on the planet's dayside compared to the nightside. This challenges the assumption of a uniform magnetic field encountered by the planetary obstacle and could lead to a larger Poynting flux directed toward the star from the dayside. Additionally, the Alfv\'en wings generated by these planets may exhibit complex cross-sectional shapes, adding further complexity to the scenario and suggesting that current estimates and scaling laws may need refinement. There may also be notable uncertainties related to the assumption that only 0.2 percent of the total energy is emitted in the Ca II K band. These uncertainties arise because current estimates are based on measurements of solar flares, where the emission is primarily driven by bremsstrahlung radiation from accelerated particles. This may not accurately reflect the situation when the power is generated by the dissipation of Alfv\'en waves in the chromosphere.

\section{Conclusions and discussions\label{sec:conclusions}}
In close-in exoplanetary systems exhibiting SPMI, Alfvén waves carry a significant amount of Poynting flux from the planet towards the star eventually leading to the formation of stellar chromospheric hotspots capable of generating observable signatures. The stellar transition region introduces a discontinuity in the Alfvén speed profile, affecting the propagation of these waves along the magnetic field lines connecting the planetary obstacle to the star. This discontinuity can reflect a portion of the wave energy back, with the reflection dependent on the wave's frequency, resulting in a frequency-dependent transmission profile of the Alfvén waves. The primary objective of this paper is to quantify the overall efficiency of energy transfer by Alfvén waves through the stellar transition region, with a specific focus on how frequency-dependent transmission affects the energy transfer process. We perform numerical studies of propagation of Alfvén waves from an exoplanet to the host star along one of the magnetic field lines connecting the two. The propagation takes place in a realistic streaming stellar wind background. The key findings of this study are summarized as follows:

\begin{itemize}
    \item Alfvén waves propagate with minimal reflection up to the transition region (TR), beyond which their transmission through the TR exhibits frequency dependence which can be reasonably quantified by the Leroy-1981 model of Alfvén wave propagation through stratified media.
    \item Quantifying the transmittance of Alfvén waves through a discontinuity such as the stellar transition region reveals that a frequency window exists, bounded by the lowest frequency wave an obstacle can physically generate at the lower end and the onset of PDI at the upper end. Within this window, energy transfer through the discontinuity by Alfvén waves is most efficient, with a significant reduction in transmittance outside this range.
    \item The efficiency of energy transfer varies with exoplanet size, stellar surface field strengths and exoplanetary polar field strengths. In all cases, the maximum efficiency achieved in the entire parameter space is approximately 70\% whereas the typical efficiency of realistic exoplanetary hot Jupiter systems is of the order of 10\%.
    \item Incorporating the efficiency factor at the transition region, the energy budget available for generating chromospheric hotspots, as predicted by the \citet{Lanza_2013} model, once again begins to approach the outstanding issue of a requirement of unrealistically high exoplanetary magnetic field strengths to account for the observed power emitted by these hotspots.
\end{itemize}

In summary, these discrepancies emphasize the need to move beyond simplistic estimations of the power budget associated with SPMI. In the application of our efficiencies to the systems observed by \citet{Cauley_2019}, we have considered that the transition region and the corona of the star were structured like the solar atmosphere. We need now to move on to parameterize adequately the structure of the transition region of cool stars to better assess the transmittance of Alfvén waves through them. Also, considering that the energy involved in SPMI can in theory generate chromospheric hot spots, the partitioning of this energy between various stellar activity tracers is still largely unknown. Detailed modeling of the heating associated with SPMI within the chromosphere would be required to assess whether, for instance, the canonical 0.2\% used in this work for Ca-II K emission is realistic or not. Such modeling efforts would also allow the assessment of energy partition into other tracers, for instance, in the visible H$\alpha$ \citep[\textit{e.g.}][]{Strugarek_2019} or other chromospheric tracers in the infrared \citep[\textit{e.g.}][]{Klein_2021}. Precise quantification of the additional efficiency factors associated with the process is essential. This includes, for example, the overall emission efficiency of Alfvén waves upon reaching the chromosphere and the efficiency factors governing their propagation to the transition region. Notably, lateral inhomogeneities along the Alfvén wings could induce phase mixing, leading to the eventual decay of the waves during propagation. However, this aspect lies beyond the scope of the present study \citep{Heyvaerts_1983}.

The small-scale topology of stellar magnetic fields can differ significantly from the large-scale magnetic flux used to characterize the star. For instance, in a Sun-like star, the global averaged flux value corresponds to approximately 1.5 G. However, local small-scale features can exhibit magnetic field strengths of the order of 1000 G, which could have potential to influence local transmittance values. A subtle point to consider is that such small-scale magnetic structures, are typically confined in height as well. Regions of very high field concentration experience pronounced superradial expansion in the chromosphere and lower corona. This expansion typically reduces the steepness of the Alfvén speed gradient, and depending on the altitude at which the expansion occurs, it can significantly enhance transmittance \citep{Similon_1992}. Therefore, in essence, relying solely on global, large-scale magnetic field strengths derived from e.g. Zeeman-Doppler Imaging (ZDI), which offers limited spatial resolution, may be insufficient to accurately characterize energy transmission in such systems.

It is important to note that large planetary magnetic fields necessarily involve large magnetospheres, which will intercept an inhomogeneous ambient medium across the day-to-night sides. As a result, a more refined approach, supported by robust analytical frameworks and validated through self-consistent 3D numerical simulations \citep[\textit{e.g.}][]{Strugarek_2015,Strugarek_2022}, is essential for a comprehensive understanding of the associated processes starting from generation of the SPMI power budget, the dissipation during the propagation through the streaming stellar wind and finally, the eventual dissipation in the chromosphere. The photosphere/chromosphere boundary itself may act as a source of Alfvén waves, which could potentially interact with SPMI waves originating from the planet. Counter-propagating waves are known to facilitate the development of Alfvénic turbulence, which could, in principle, influence the overall transmittance. However, investigating the effects of nonlinear interactions between counter-propagating waves on transmittance profiles represents a complex and extensive area of study. While this line of inquiry is beyond the scope of the current work, it presents a promising avenue for future research. Ultimately, these findings reveal that there is yet much to be explored in the complex physics underlying SPMI.

\begin{acknowledgements}
     We acknowledge funding from the Programme National de Planétologie (INSU/PNP). A.S. acknowledges funding from the European Union’s Horizon-2020 research and innovation programme (grant agreement no. 776403 ExoplANETS-A) and the PLATO/CNES grant at CEA/IRFU/DAp, and the European Research Council project ExoMagnets (grant agreement no. 101125367). This project was provided with computing HPC and storage resources by GENCI at TGCC thanks to the grant 2024-A0160410133 on the supercomputer Joliot Curie's SKL and Rome partition. AP acknowledges funding from the MERAC foundation.
\end{acknowledgements}

%
%

\bibliography{biblio}{} 
\bibliographystyle{aa}
\end{document}